\documentclass[11pt]{article}

\usepackage{latexsym}
\usepackage{amsmath}
\usepackage[dvips]{epsfig}
\usepackage{graphicx}

\def\beq{\begin{eqnarray}}
\def\eeq{\end{eqnarray}}

\def\Tu{\hat T}
\def\Td{\check T}
\def\Tb{\bar T}

\def\X{{\cal X}}
\def\Y{{\cal Y}}

\def\MOD{\mbox{mod}}

\def\T{^{\scriptscriptstyle T}}

\begin{document}

\fontsize{11}{14.5pt}\selectfont

\begin{center}

{\small 
Technical Report No.\ 1011, Department of Statistics, University of Toronto}

\vspace*{1in}

\begin{center} \LARGE \bf 
 MCMC Using Ensembles of States for \\[4pt] 
 Problems with Fast and Slow Variables such as \\[4pt] 
 Gaussian Process Regression
\end{center}

\vspace{10pt}

{\large Radford M. Neal \\[4pt]
  \normalsize Dept.\ of Statistics and Dept.\ of Computer Science \\
  University of Toronto \\[4pt]
  \texttt{http://www.cs.utoronto.ca/$\sim$radford/}\\
  \texttt{radford@stat.utoronto.ca}\\[4pt]
  31 December 2010}
 
\end{center}

\vspace{10pt}

\noindent {\bf Abstract.}  I introduce a Markov chain Monte Carlo
(MCMC) scheme in which sampling from a distribution with density
$\pi(x)$ is done using updates operating on an ``ensemble'' of states.
The current state $x$ is first stochastically mapped to an ensemble,
$(x^{(1)},\ldots,x^{(K)})$.  This ensemble is then updated using MCMC
updates that leave invariant a suitable ensemble density,
$\rho(x^{(1)},\ldots,x^{(K)})$, defined in terms of $\pi(x^{(i)})$ for
$i\,=\,1,\ldots,K$.  Finally a single state is stochastically selected
from the ensemble after these updates.  Such ensemble MCMC updates can
be useful when characteristics of $\pi$ and the ensemble permit
$\pi(x^{(i)})$ for all $i \in \{1,\ldots,K\}$, to be computed in less
than $K$ times the amount of computation time needed to compute
$\pi(x)$ for a single $x$.  One common situation of this type is when
changes to some ``fast'' variables allow for quick re-computation of
the density, whereas changes to other ``slow'' variables do not.
Gaussian process regression models are an example of this sort of
problem, with an overall scaling factor for covariances and the noise
variance being fast variables.  I show that ensemble MCMC for Gaussian
process regression models can indeed substantially improve sampling
performance.  Finally, I discuss other possible applications of
ensemble MCMC, and its relationship to the ``multiple-try
Metropolis'' method of Liu, Liang, and Wong and the ``multiset
sampler'' of Leman, Chen, and Lavine.

\subsection*{Introduction}\vspace*{-8pt}

In this paper, I introduce a class of Markov chain Monte Carlo methods
that utilize a state space that is the $K$-fold Cartesian product of
the space of interest --- ie, although our interest is in sampling for
$x$ in the space $\X$, we use MCMC updates that operate on
$(x^{(1)},\ldots, x^{(K)})$ in the space $\Y \,=\, \X^K$.

Several such methods have previously been proposed --- for example,
Adaptive Directive Sampling (Gilks, Roberts, and George, 1994) and
Parallel Tempering (Geyer, 1991; Earl and Deem, 2005).  The ``ensemble
MCMC'' methods I introduce here differ from these methods in two
fundamental respects.  First, use of the space $\Y\,=\,\X^K$ is
temporary --- after some number of updates on $\Y$, one can switch
back to the space $\X$, perform updates on that space, then switch
back to $\Y$, etc.  (Of course, one might choose to always remain in
the $\Y$ space, but the option of switching back and forth exists.)
Secondly, the invariant density for an ensemble state
$(x^{(1)},\ldots,x^{(K)})$ in the $\Y$ space is proportional to the
\textit{sum} of the probabilities of the individual $x^{(k)}$ (or more
generally a weighted sum), whereas in the previous methods mentioned
above, the density is a \textit{product} of densities for each
$x^{(k)}$.  As a consequence, the properties of the ensemble methods
described here are quite different from those of the methods mentioned
above.

I expect that use of an ensemble MCMC method will be advantageous only
when, for the particular distributions being sampled from, and the
particular ensembles chosen, a computational short-cut exists that
allows computation of the density for all of $x^{(1)},\ldots,x^{(K)}$
with less than $K$ times the effort needed to compute the density for
a single state in $\X$.  Without this computational advantage, it is
hard to see how performing updates on $\X^K$ could be beneficial.

In this paper, I mostly discuss one particular context where such a
computational short-cut exists --- when the distribution being sampled
has the characteristic that after changes to only a subset of ``fast''
variables it is possible to re-compute the density in much less time
than is needed when other ``slow'' variables change.  By using
ensembles of states in which the slow variables have the same values
in all ensemble members, the ensemble density can be quickly computed.
In the limit as the size of the ensemble grows, it is possible using
such a method to approach the ideal of updating the slow variables
based on their marginal distribution, integrating over the fast
variables.

I apply these fast-slow ensemble MCMC methods to Gaussian process
regression models (Rasmussen and Williams 2006; Neal 1999) that have a
covariance function with unknown parameters, which in a fully Bayesian
treatment need to be sampled using MCMC.  The computations for such
models require operations on the covariance matrix whose time grows in
proportion to $n^3$, where $n$ is the number of observations.  If
these computations are done using the Cholesky decomposition of the
covariance matrix, a change to only the overall scale of the
covariance function does not require recomputation of the Cholesky
decomposition; hence this scale factor can be a fast variable.  If
computations are done by finding the eigenvectors and eigenvalues of
the covariance function, both an overall scale factor and the noise
variance can be fast variables.  I show that ensemble MCMC with either
of these approaches can improve over MCMC on the original state space.

I conclude by discussing other possible applications of ensemble MCMC,
and its relationship to the ``multiple-try Metropolis'' method of Liu,
Liang, and Wong (2000), and to the ``multiset sampler'' of Leman,
Chen, and Lavine (2009).

\subsection*{MCMC with an ensemble of states}\vspace*{-8pt}

Here, I introduce the idea of using an ensemble of states in general,
using the device of stochastically mapping from the space of interest
to another space.  I also discuss several ways of defining an ensemble.

I will suppose that we wish to sample from a distribution on $\X$ with
probability density or mass function $\pi(x)$.  The MCMC approach is
to define a Markov chain that has $\pi$ as an invariant distribution.
Provided this chain is ergodic, simulating the chain from any initial
state for a suitably large number of transitions will produce a state
whose distribution is close to $\pi$.  To implement this strategy, we
need to define a transition probability, $T(x'|x)$, for the chain to
move to state $x'$ if it is currently in state $x$.  Updates made
according to this transition probability must leave $\pi$ invariant:
\beq
  \int \pi(x)\, T(x'|x)\, dx & = & \pi(x')
\eeq

\paragraph{Stochastically mapping to an ensemble and back.}  
An MCMC update that uses an ensemble of $K$ states can be viewed as
probabilistically mapping from the original space, $\X$, to a new
space, $\Y = \X^K$, performing updates on this new space, and then
mapping back to the original space.  We can formalize this ``temporary
mapping'' strategy, which has many other applications (Neal 2006;
Neal 2010, Section 5.4), as follows.  We define a transition distribution,
$T(x'|x)$, on the space $\X$ as the composition of three other
stochastic mappings, $\Tu$, $\Tb$, and $\Td$:
\beq x \
\
 \begin{array}{cc} \scriptstyle\Tu \\[-7pt] \longrightarrow
 \\[6pt]\end{array}\ \
y \ \ 
 \begin{array}{cc} \scriptstyle\Tb \\[-7pt] \longrightarrow
 \\[6pt]\end{array}\ \
y'\ \
 \begin{array}{cc} \scriptstyle\Td \\[-7pt] \longrightarrow
 \\[6pt]\end{array}\ \
x'
\eeq
That is, starting from the current state, $x$, we obtain a value in the 
temporary space by sampling
from  $\Tu(y|x)$.  The target distribution for $y \in \Y$ has 
probability/density function $\rho(y)$.  We require that
\beq
  \int \pi(x)\, \Tu(y|x)\, dx & = & \rho(y)\label{eq-upinv}
\eeq
$\Tb(y'|y)$ defines a transition on the temporary space that leaves $\rho$
invariant:
\beq
  \int \rho(y)\, \Tb(y'|y)\, dy & = & \rho(y')\label{eq-barinv}
\eeq
Finally, $\Td$ gets us back to the original space.  It must satisfy
\beq
  \int \rho(y')\, \Td(x'|y')\, dy' & = & \pi(x')\label{eq-downinv}
\eeq
The above conditions imply that the overall transition, $T(x'|x)$, will leave
$\pi$ invariant, and so can be used for Markov sampling of $\pi$.

In ensemble MCMC, where $\Y = \X^K$, with the ensemble state 
written as $y = (x^{(1)},\ldots,x^{(K)})$,
the stochastic mapping $\Tu$, from $\X$ to $\Y$, is defined in terms of
an \textit{ensemble base measure}, a distribution with probability density or
mass function $\zeta(x^{(1)},\ldots,x^{(K)})$, 
as follows:
\beq
  \Tu(x^{(1)},\ldots,x^{(K)}\,|\,x) & = &
  {1 \over K} \sum_{k=1}^K \zeta_{-k|k}(x^{(-k)}\,|\,x)\,\delta_x(x^{(k)})
  \label{eq-ensup}
\eeq
Here, $\delta_x(\cdot)$ is the distribution with mass concentrated at $x$.
The notation $x^{(-k)}$ refers to all components of the ensemble state
other than the $k$'th.  The probability density or mass function for the
conditional distribution of components other than the $k$'th given a
value for the $k$'th component is $\zeta_{-k|k}(x^{(-k)}|x^{(k)}) =
\zeta(x^{(1)},\ldots,x^{(K)})\,/\,\zeta_k(x^{(k)})$, where $\zeta_k$
is the marginal density or mass function for the $k$'th component, 
derived from the joint distribution $\zeta$.  We assume that $\zeta_k(x)$
is non-zero if $\pi(x)$ is non-zero.

Algorithmically, $\Tu$ creates a state in $\Y$ by choosing an index,
$k$, from $1$ to $K$, uniformly at random, setting $x^{(k)}$ to the
current state, $x$, and finally randomly generating all $x^{(j)}$ for
$j \ne k$ from their conditional distribution under $\zeta$ given that
$x^{(k)}=x$.   Note that $\Tu$ depend only on $\zeta$, not on $\pi$.

\vspace{-4pt}
\noindent\rule{6.7in}{0.5pt}\\
\textit{A simple example:}\ \ As a simple (but not useful) illustration,
let $\X = [0,2\pi)$ and set $K=2$.  
Choose the ensemble base measure, $\zeta$, to be uniform on
pairs of points with $x^{(2)}\, =\, x^{(1)}+\omega\ 
(\MOD\ 2\pi)$,
for some constant $\omega \in [0,2\pi)$.  (Ie, the ensemble
base measure is uniform over pairs of points on a unit circle with the 
second point at an angle $\omega$ counterclockwise from the first.)
Then $\zeta_{-1|1}(\,\cdot\,|x^{(1)}) \,
=\, \delta_{x^{(1)}+\omega \ (\MOD\ 2\pi)}(\,\cdot\,)$
and $\zeta_{-2|2}(\,\cdot\,|x^{(2)}) \,
=\, \delta_{x^{(1)}-\omega \ (\MOD\ 2\pi)}(\,\cdot\,)$.  The
algorithm for the map $\Tu$ is\vspace{8pt}\\
\mbox{~~~~~~~}1) pick $k$ uniformly from $\{1,2\}$ \\
\mbox{~~~~~~~}2) set $x^{(k)} = x$ \\
\mbox{~~~~~~~}3) if $k=1$, set $x^{(2)}=x+\omega\ (\MOD\ 2\pi)$;\ \
                 if $k=2$, set $x^{(1)}=x-\omega\ (\MOD\ 2\pi)$. \\
\rule{6.7in}{0.5pt}

Having defined $\Tu$ as in equation~\eqref{eq-ensup}, the 
density function, $\rho$, for the \textit{ensemble distribution} that will
make condition \eqref{eq-upinv} hold can be derived as follows:
\beq
  \rho(x^{(1)},\ldots,x^{(K)}) & = &
  \int \pi(x)\, \Tu(x^{(1)},\ldots,x^{(K)}\,|\,x)\,dx \\[6pt]
  & = & 
  {1 \over K} \sum_{k=1}^K \zeta_{-k|k}(x^{(-k)}\,|\,x^{(k)})\,\pi(x^{(k)})
  \label{eq-rhodef1}\\[6pt]
  & = & {1 \over K} \sum_{k=1}^K 
  {\zeta(x^{(1)},\ldots,x^{(K)}) \over \zeta_k(x^{(k)})} \,\pi(x^{(k)})
  \\[6pt]
  & = & \zeta(x^{(1)},\ldots,x^{(K)})\, {1 \over K} \sum_{k=1}^K 
  {\pi(x^{(k)}) \over \zeta_k(x^{(k)})}\label{eq-rhodef2}
\eeq
So we see that the density function of $\rho$ with respect
to the base measure $\zeta$ is sometimes simply proportional to the
sum of $\pi$ for all the ensemble members (when the $\zeta_k$
are all the same uniform distribution), and more generally is proportional to
the sum of ratios of $\pi$ and $\zeta_k$.

\vspace{-4pt}
\noindent\rule{6.7in}{0.5pt}\\
\textit{The simple example continued:}\ \ 
$\zeta_1$ and $\zeta_2$ are both uniform
over $[0,2\pi)$.  For pairs in $\X^2$ satisfying the constraint that
$x^{(2)}\,=\,x^{(1)}+\omega\ (\MOD 2\pi)$, the ensemble density follows 
the proportionality:
\beq
   \rho(x^{(1)},x^{(2)}) & \propto & \pi(x^{(1)}) \,+\, \pi(x^{(2)})
\eeq
(The probability under $\rho$ of a pair that violates the constraint 
is of course zero.) \\
\rule{6.7in}{0.5pt}

We can let $\Tb$ be any sequence of Markov updates for
$y\,=\,(x^{(1)},\ldots,x^{(K)})$ that leave $\rho$ invariant
(condition~\eqref{eq-barinv}).  We denote the result of applying $\Tb$
to $y$ by $y'$.

\vspace{-4pt}
\noindent\rule{6.7in}{0.5pt}\\
\textit{The simple example continued:}\ \ We might define $\Tb$ to
be a sequence of some pre-defined number of Metropolis updates (Metropolis,
\textit{et al} 1953), using a random-walk proposal, which from state 
$(x^{(1)},\,x^{(2)})$
is $(x_*^{(1)},\,x_*^{(2)})\ =\ 
(x^{(1)}+\upsilon\ (\MOD 2\pi),\,x^{(2)}+\upsilon\ (\MOD 2\pi))$,
with $\upsilon$ being drawn from a distribution symmetrical
around zero, such as uniform on $(-\pi/10,\,
+\pi/10)$.  Such a proposal is accepted with probability $\min [1,\
(\pi(x^{(1)}_*)+\pi(x^{(2)}_*))\,/\,(\pi(x^{(1)})+\pi(x^{(2)})]$. \\
\rule{6.7in}{0.5pt}

To return to a single state, $x'$, from the ensemble state $y' =
(x^{(1)},\ldots,x^{(K)})$, we set $x'$ to one of the $x^{(k)}$,
which we select randomly with probabilities proportional to
$\pi(x^{(k)})\,/\,\zeta_k(x^{(k)})$.  We can see that this $\Td$ satisfies
condition~\eqref{eq-downinv} as follows:
\beq
  \lefteqn{\int \rho(y')\, \Td(x'|y')\, dy'}\ \ \ \nonumber\\[3pt]
  & = & \!\!\!
  \int\! \left[ \zeta(x^{(1)},\ldots,x^{(K)})\, {1 \over K} \sum_{k=1}^K 
  {\pi(x^{(k)}) \over \zeta_k(x^{(k)})}\right] \!
  \left[ 
    \sum_{j=1}^K \delta_{x^{(j)}}(x')\,{\pi(x^{(j)}) \over \zeta_j(x^{(j)})}
    \,\Big/ \sum_{k=1}^K {\pi(x^{(k)}) \over \zeta_k(x^{(k)})}
  \right] dx^{(1)}\cdots dx^{(K)}\ \ \ \ \ \ \\[3pt]
  & = &
  { 1 \over K } \sum_{j=1}^K \int \delta_{x^{(j)}}(x')\, \pi(x^{(j)}) \,
    {\zeta(x^{(1)},\ldots,x^{(K)}) \over \zeta_j(x^{(j)})} 
    \,dx^{(1)}\cdots dx^{(K)}\ \ \ \ \\[3pt]
  & = &
  { 1 \over K } \sum_{j=1}^K \int \pi(x')\, \zeta_{-j|j}(x^{(-j)}|x')\,
    dx^{(-j)}\ \ \ \ \\[3pt]
  & = &   { 1 \over K } \sum_{j=1}^K \pi(x')
          \ \ =\ \ \pi(x')
\eeq

\vspace{-4pt}
\noindent\rule{6.7in}{0.5pt}\\
\textit{The simple example continued:}\ \ Since $\zeta_1$ and
$\zeta_2$ are uniform over $[0,2\pi)$, $\Td$ simply picks either
$x^{(1)}$ or $x^{(2)}$ with probabilities proportional to $\pi(x^{(1)})$
and $\pi(x^{(2)})$. \\
\rule{6.7in}{0.5pt}

\paragraph{Some possible ensembles.} The ensemble base measure,
$\zeta$, can be defined in many ways, some of which I will discuss
here.  Note that the usefulness of these ensembles depends on whether
they produce any computational advantage for the distribution being
sampled.  This will be discussed in the next section for problems with
fast and slow variables, where variations on some of the ensembles
discussed here will be used.

One possibility is that $x^{(1)},\ldots,x^{(K)}$ are independent and
identically distributed under $\zeta$, so that
$\zeta(x^{(1)},\ldots,x^{(K)})\,=\,\zeta(x^{(1)})\cdots\zeta(x^{(K)})$,
where $\zeta(x)$ is the marginal distribution of all the $x^{(k)}$.
In the conditional distribution $\zeta_{-k|k}$, the components other
than $x^{(k)}$ will also be independent, with distribution
$\zeta(x^{(k)})$ the same as their marginal distribution.  With this
ensemble base measure, the ordering of states in the ensemble is
irrelevant, and the mapping, $\Tu$, from a single state to an ensemble
consists of combining the current state with $K-1$ states sampled from
the marginal $\zeta$.  The mapping, $\Td$, from the ensemble to a single
state consists of randomly selecting a state from
$x^{(1)},\ldots,x^{(k)}$ with probabilities proportional to
$\pi(x^{(k)})/\zeta(x^{(k)})$. 

Rather than the $x^{(k)}$ being independent in the ensemble base
measure, they might just be exchangeable.  This can be expressed as
the $x^{(i)}$ being independent given some paramater $\theta$, which
has some ``prior'', $\zeta(\theta)$ (which is unrelated to the
real prior for a Bayesian inference problem).  In other words,
\beq
  \zeta(x^{(1)},\ldots,x^{(K)}) & = &
  \int \zeta(\theta) \prod_{k=1}^K\, \zeta(x^{(k)}|\theta)\, d\theta
\eeq
The mapping $\Tu$ can be implemented by sampling $\theta$ from the
``posterior'' density $\zeta(\theta\,|\,x^{(k)}=x) \propto \zeta(\theta)\,
\zeta(x^{(k)}=x\,|\,\theta)$, for $k$ randomly chosen from $1,\ldots,K$
(though due to exchangeability, this random choice of $k$ isn't really 
necessary) and then sampling $x^{(j)}$ for $j \ne k$ independently
from $\zeta(x^{(j)}|\theta)$ (which is the same for all $j$).  The
marginal distributions for the $x^{(k)}$ in the ensemble base measure,
needed to compute $\rho$ and to map from an ensemble to a single state,
are of course all the same when this measure is exchangeable.

Another possibility is for the ensemble base measure on
$x^{(1)},\ldots,x^{(K)}$ to be defined by a stationary Markov chain,
which again leads to the $x^{(k)}$ all having the same marginal
distribution, $\zeta(x)$.  For example, if $\X$ is the reals, we might
let $\zeta(x)\,=\,N(x;\, 0,\tau^2)$, where $N(x;\, \mu,\tau^2)$ is the
normal density function, and $\zeta_{k+1|k}(x^{(k+1)}|x^{(k)})
\,=\,N(x^{(k+1)};\, x^{(k)}\sqrt{1-1/\tau^2},\, 1)$ for
$k=1,\ldots,K\!-\!1$.  Going from a single state to an ensemble is
done by randomly selecting a position, $k$, for the current state, and
then simulating the Markov chain forward from $x^{(k)}$ to $x^{(K)}$
and backward from $x^{(k)}$ to $x^{(1)}$ (which will be the same as
forward simulation when the chain is reversible, as in the example
here).  The ensemble density is given by equation~\eqref{eq-rhodef2},
with all the marginal densities $\zeta_k(x)$ equal to $\zeta(x)$.  We
return from an ensemble to a single state by selecting from
$x^{(1)},\ldots,x^{(k)}$ with probabilities proportional to
$\pi(x^{(k)})/\zeta(x^{(k)})$.

Ensemble base measures can also be defined using
\textit{constellations} of points.  Let $x_*^{(1)},\ldots,x_*^{(K)}$
be some set of $K$ points in $\X$.  We let $\zeta$ be the distribution
obtained by shifting all these points by an amount, $\theta$, chosen from some
distribution on $\X$, with density $\zeta(\theta)$ that is nowhere zero,
so that
\beq
  \zeta(x^{(1)},\ldots,x^{(K)}) & = & 
  \int \zeta(\theta) \prod_{k=1}^K \delta_{x_*^{(k)}+\theta\ }(x^{(k)})\,d\theta
\eeq
The conditional distributions $\zeta_{-k|k}$ do not depend on
$\zeta(\theta)$ --- they are degenerate distributions in which the
other constellation points are determined by $x^{(k)}$.  We therefore
move from a single point, $x$, to a constellation of points by
selecting $k$ at random from $1,\ldots,K$, setting $x^{(k)}$ to $x$,
and setting $x^{(j)}$ for $j \ne k$ to $x_*^{(j)}+x-x_*^{(k)}$.  The
ensemble density, $\rho(x^{(1)},\ldots,x^{(K)})$, will also not depend
on $\zeta(\theta)$, as can be seen from equation~\eqref{eq-rhodef1})
--- it will be proportional to the sum of $\pi(x^{(k)})$ for ensembles
that have the shape of the constellation, and be zero for those that do
not.  The marginal densities, $\zeta_k(x^{(k)})$, will be the same for
all constellation points, since $\theta = x^{(k)}-x_*^{(k)}$ will be
the same for all $k$. (Note that this is not the same as the marginal
density functions being the same for all $k$.)  Hence moving to a
single point from a constellation is done by choosing a point,
$x^{(k)}$, from the constellation with probabilities proportional to
$\pi(x^{(k)})$.

The simple example used in the previous section can be seen as a
constellation of two points with $x^{(1)}_*=0$, $x^{(2)}_*=\omega$,
$\zeta(\theta)$ uniform over $[0,2\pi)$, and addition done modulo $2\pi$.

In the presentation above, the ensemble base measure $\zeta$ is
assumed to be a proper probability distribution, but $\zeta$ can
instead be an improper limit of proper distributions, as long as the
conditional distributions and ratios of marginal distributions that
are needed have suitable limits.  In particular, the conditional
distributions $\zeta_{-k|k}$ used to define $\Tu$ in
equation~\eqref{eq-ensup} must reach limits that are proper
distributions; this will also ensure that $\rho$ is well defined (see
equation~\eqref{eq-rhodef1}).  The ratios
$\zeta_j(x^{(j)})\,/\,\zeta_k(x^{(k)})$ must also reach limits,
so that $\Td$ will be well defined.

We can obviously let $\zeta(\theta)$ become improper when defining a
constellation ensemble, since we have seen that the choice of this
density has no effect.  For exchangeable ensembles, we can let
$\zeta(\theta)$ be improper as long as the ``posterior'',
$\zeta(\theta\,|\,x^{(k)}=x)$, is proper.  We can also let $\tau$ go
to infinity in the Markov ensemble base measure defined above.  We
will then have $\zeta_{k+1|k}(x^{(k+1)}|x^{(k)})
\,=\,N(x^{(k+1)};\,x^{(k)},1)$, so the conditional distributions
$\zeta_{-k|k}$ are simple random walks in both directions from
$x^{(k)}$.  Since the marginal distributions for the $x^{(k)}$ are all
normal with mean zero and variance $\tau^2$, for any $x^{(j)}$ and
$x^{(k)}$, the ratio $\zeta(x^{(j)})\,/\,\zeta(x^{(k)})$ approaches
one as $\tau$ goes to infinity.  (Note also that since the limit of
$\zeta_{-k|k}$ is proper, the relevant values of $x^{(k)}$ will not
diverge, so uniform convergence of these ratios is not necessary.)

\subsection*{Ensemble MCMC for problems with ``fast'' and ``slow'' variables}
\vspace*{-8pt}

Suppose that we wish to sample from a distribution on a space that can
be written as a Cartesian product, $\X=\X_1\times\X_2$, with elements
written as $x=(x_1,x_2)$.  Suppose also that the probability density
or mass function, $\pi(x_1,x_2)$, is such that re-computing the
density after a change to $x_2$ is much faster than recomputing the
density after a change to $x_1$.  That is, if we have computed
$\pi(x_1,x_2)$, and saved suitable intermediate results from this
computation, we can quickly compute $\pi(x_1,x_2')$ for any $x_2'$,
but there is no short-cut for computing $\pi(x_1',x_2')$ for $x_1' \ne
x_1$.  In general, $x_1$ and $x_2$ might both be multidimensional.  I
refer to $x_1$ as the ``slow'' variables, and $x_2$ as the ``fast''
variables.

I first encountered this class of problems in connection with models
for the Cosmic Microwave Background (CMB) radiation (Lewis and Bridle,
2002).  This is an example of data modelled using a complex
simulation, in this case, of the early universe.  In such problems,
the slow variables, $x_1$, are the unknown parameters of the
simulation, which are being fit to observed data.  When new values for
these parameters are considered, as for a Metropolis proposal,
computing $\pi$ with these new values will require re-running the
simulation, which we assume takes a large amount of computation time.
Some fairly simple model relates the output of the simulation to the
observed data, with parameters $x_2$ (for example, the noise level of
the observations).  When a new value for $x_2$ is considered in
conjunction with a value for $x_1$ that was previously considered, the
simulation does not need to be re-run, assuming the output from the
previous run was saved.  Instead, computing $\pi$ with this new $x_2$
and the old $x_1$ requires only some much faster calculation involving
the observation model.

Fast and slow variables arise in many other contexts as well.  For
example, in many Bayesian models, the posterior density for a set of
low-level parameters involves a large number of data items, which
cannot be summarized in a few sufficient statistics, while a small
number of hyperparameters have densities that depend (directly) only
on the low-level parameters.  The low-level parameters in such a
situation will be ``slow'', since when they are changed, all the data
items must be looked at again, but the hyperparameters may be
``fast'', since when they change, but the low-level parameters stay
the same, there is no need to look at the data.

Later in this paper, I will consider applying methods for fast and
slow variables to the more specialized problem of sampling from the
posterior distribution of the parameters of a Gaussian process
regression model, in which an overall scale factor for the covariance
function and the noise variance can be regarded as fast variables.

\paragraph{Previous methods for problems with fast and slow variables.}

A simple way of exploiting fast variables, used by Lewis and Bridle
(2002), is to perform extra Metropolis updates that change only the
fast updates, along with less frequent updates that change both fast
and slow variables.  Lewis and Bridle found that this is an
improvement over always performing updates for all variables.
Similarly, when using a Metropolis method that updates only one
variable at a time, one could perform more updates for the fast
variables than for the slow variables.  I will refer to these as
``extra update Metropolis'' methods.

Another simple fast-slow method I devised (that has been found useful
for the CMB modeling problem) is to randomly choose a displacement in
the slow space, $\delta_1 \in \X_1$, and then perform some
pre-determined number of Metropolis updates in which the proposal from
state $(x_1,x_2)$ is $(x_1\pm\delta_1,x_2')$, where the sign before
$\delta_1$ is chosen randomly, and $x_2'$ is chosen from some suitable
proposal distribution, perhaps depending on $x_1$, $x_2$, and the
chosen sign for $\delta_1$.  During such a sequence of ``random grid
Metropolis'' updates, all proposed and accepted states will have slow
variables of the form $x_1+i\delta_1$, where $x_1$ is the initial
value, and $i$ is an integer.  If intermediate results of computations
with all these values for the slow variables are saved, the number of
slow computations may be much less than the number of Metropolis
proposals, since the same slow variables may be proposed many times in
conjunction with different values for the fast variables.

If recomputation after a change to the fast variables is very much
faster than after a change to the slow variables, we might hope to
find an MCMC method that samples nearly as efficiently as would be
possible if we could efficiently compute the marginal distribution for
just the slow variables, $\pi(x_1) = \int \pi(x_1,x_2)\, dx_2$, since
we could approximate this integral using many values of $x_2$.  I have
devised a ``dragging Metropolis'' method (Neal, 2004) that can be seen
as attempting to achieve this.  The ensemble MCMC methods I describe
next can also be viewed in this way.

\paragraph{Applying ensemble MCMC to problems with fast and slow variables.}
Ensemble MCMC can be applied to this problem using ensembles in which
all states have the same value for the slow variables, $x_1$.
Computation of the ensemble probability density
(equation~\eqref{eq-rhodef2}) can then take advantage of quick
computation for multiple values of the fast variables.  Many sampling
schemes of this sort are possible, distinguished by the ensemble base
measure used, and by the updates that are performed on the ensemble
(ie, by the transition $\Tb$). 

For all the schemes I have looked at, the fast and slow variables
are independent in the base ensemble measure used, and the slow
variables have the same value in all members of the ensemble.  The
ensemble base measure therefore has the following form:
\beq
   \zeta(x^{(1)},\ldots,x^{(K)}) & = &
     \Big[ \omega(x^{(1)}_1) \prod_{k=2}^K \delta_{x^{(1)}_1}(x^{(k)}_1) \Big]
     \ \xi(x^{(1)}_2,\ldots,x^{(K)}_2)
\eeq
Here, $\omega(x_1)$ is some density for the slow variables --- the choice of
$\omega$ will turn out not to matter, as long as it is nowhere zero.  
That the slow variables have the same values in all ensemble members
is expressed by the factors of $\delta_{x^{(1)}_1}(x^{(k)}_1)$.
The joint distribution of the fast variables in the $K$ members of the
ensemble is given by the density function $\xi$.  If we let $\xi_k$
be the marginal distribution for the $k$'th member of the ensemble
under $\xi$, we can write
the ensemble density from equation~\eqref{eq-rhodef2} as follows:
\beq
  \rho(x^{(1)},\ldots,x^{(K)}) & = &
  \zeta(x^{(1)},\ldots,x^{(K)})\, {1 \over K} \sum_{k=1}^K 
  {\pi(x^{(k)}) \over \zeta_k(x^{(k)})} \\[6pt]
  & = &
  \Big[ \prod_{k=2}^K \delta_{x^{(1)}_1}(x^{(k)}_1) \Big]\,
  \xi(x^{(1)}_2,\ldots,x^{(K)}_2)\,
  {1 \over K} \sum_{k=1}^K {\pi(x^{(k)}) \over \xi_k(x^{(k)}_2)}\ \ \
  \label{eq-fs-rho1}
\eeq

Possible choices for $\xi$ include those analogous to some of the general
choices for $\zeta$ discussed in the previous section, in particular the
following:\vspace{-6pt}
\begin{description}
\item[\bf ~~~independent]~\\[6pt] 
          Let $x^{(1)}_2,\ldots,x^{(K)}_2$ be independent under $\xi$, 
          with all having the same marginal distribution.  For example,
          if the fast variables are parameters in a Bayesian model,
          we might use their prior distribution.
\item[\bf ~~~exchangeable]~\\[6pt] 
          Let $x^{(1)}_2,\ldots,x^{(K)}_2$ be exchangeable under $\xi$.  
          If $x_2$ is scalar, 
          we might let $x^{(k)}_2|\theta \sim N(\theta,\tau^2)$ and
          $\theta \sim N(0,\psi^2)$. We could then let $\psi$ 
          go to infinity, so that $\xi$ is improper, and all $\xi_k$
          are the same.
\item[\bf ~~~grid points]~\\[6pt] 
          Let $x^{(k)}_2 \,=\, x_*^{(k)}\,+\, \theta$ for $k=1,\ldots,K$.  
          Here, if $x_2$ has
          dimension $D$, $\theta$ is a $D$-dimensional offset that is applied
          to the set of points $x_*^{(k)}$.  This is an example 
          of a constellation, as discussed in the previous section.  
          One possibility is a rectangular grid, with $K=m^D$ for some integer 
          $m$, and with grid points spaced a distance $d_j$ in dimension $j$, so
          that the extent of the grid in dimension $j$ is 
          $(m\!-\!1)d_j$.\vspace{-6pt}
\end{description}

The ensemble update, $\Tb$, could be done in many ways, as long as the
update leaves $\rho$ invariant.  A simple and general option is to
perform some number of Metropolis-Hastings updates (Metropolis,
\textit{et al} 1953; Hastings 1970), in which we propose to change the
current ensemble state, $y = (x^{(1)},\ldots,x^{(K)})$, to a proposed
state, $y^*$, drawn according to some proposal density $g(y^*|y)$.  We
accept the proposed $y^*$ as the new state with probability
$\min[\,1,\ \rho(y^*)\,g(y|y^*) \,/\, \rho(y)\,g(y^*|y)\,]$, where
$\rho$ is defined by equation~\eqref{eq-fs-rho1}.  A technical issue
arises when $\X$ is continuous --- since $x^{(k)}_1$ must be the same
for all members of the ensemble, the probability densities will be
infinite (eg, due to the delta functions in
equation~\eqref{eq-fs-rho1}).  We can avoid this by looking at
densities for only $x_1^{(1)}$ and $x^{(1)}_2,\ldots, x^{(K)}_2$,
since $x_1^{(2)},\ldots,x_1^{(K)}$ are just equal to $x_1^{(1)}$.
(This assumes that we never propose states violating this constraint.)
A similar issue arises when using an ensemble of grid points, as
defined above, in which the $x^{(k)}_2$ are constrained to a grid ---
here also, we can simply leave out the infinite factors in the density
that enforce the deterministic constraints, assuming that our
proposals never violate them.

\begin{figure}[b]

\centerline{\psfig{file=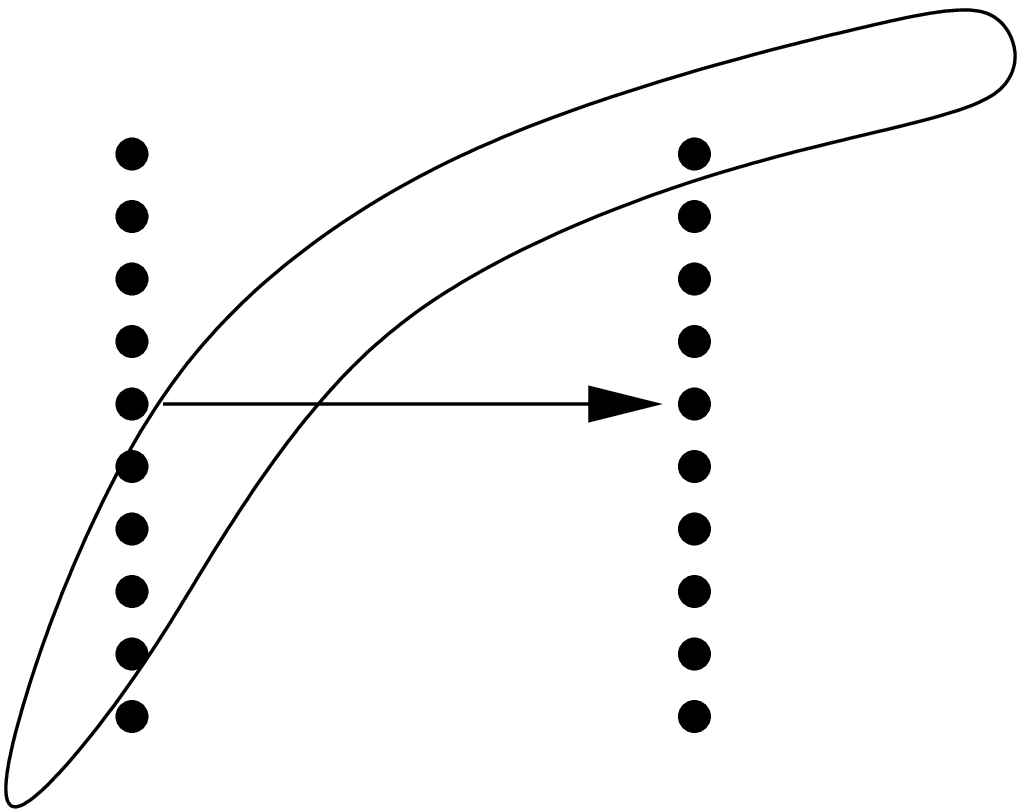,height=1.9in}%
\hspace{50pt}\psfig{file=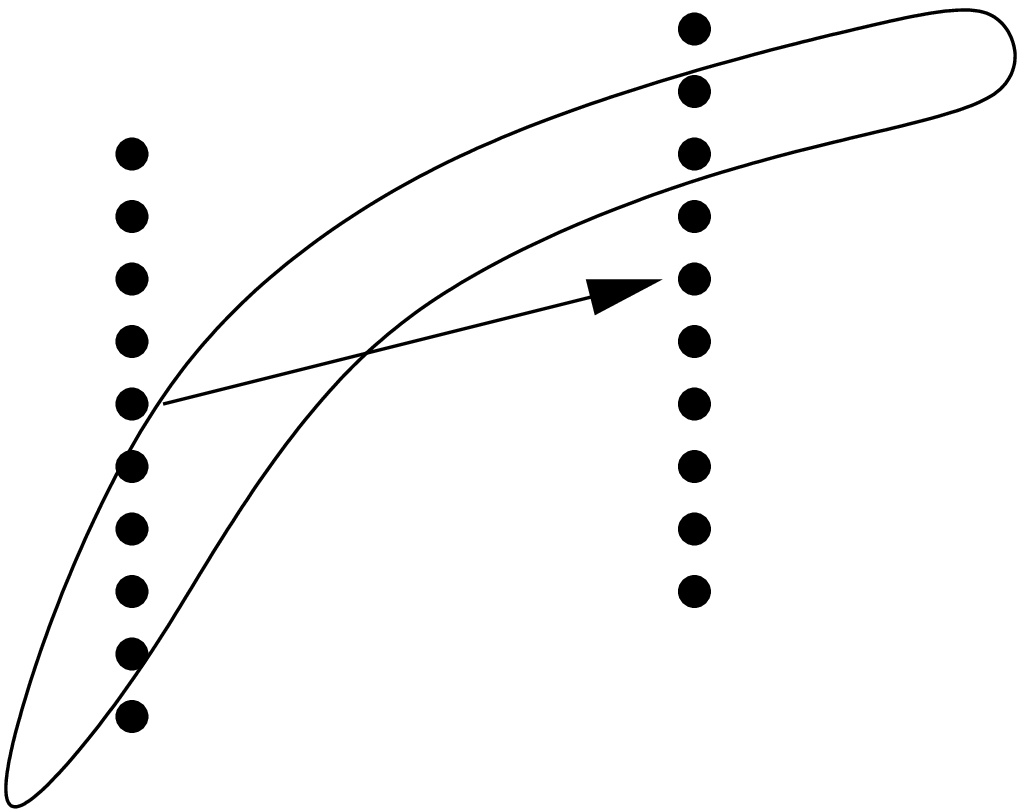,height=1.9in}}

\caption[]{Illustration of ensemble updates with with one slow
variable (horizontal) and one fast variable (vertical).  The
distribution is uniform over the outlined region.  A grid ensemble
with $K=10$ is used.  On the left, a move is proposed by changing only
the slow variable; it will be accepted with probability $1/4$, since
four members of the current ensemble have non-zero probability,
compared with only one in the proposed ensemble.  On the right, a move
is proposed by changing the slow variable and also shifting the grid
ensemble; it will be accepted with probability $2/4$.  (Of course, a
less favourable shift in the grid ensemble might have led to zero
probability of acceptance.) }\label{fig-ens-up}

\end{figure}

I will consider two classes of Metropolis updates for the ensemble
state, in which symmetry of the proposal makes $g(y|y^*)/g(y^*|y)$
equal to one, so the acceptance probability is $\min[\,1,\
\rho(y^*)/\rho(y)\,]$:\vspace*{-6pt}
\begin{description}
\item[\bf ~~~fast variables fixed]~\\[6pt] 
      Propose a new value for $x_1$ in all ensemble members, by adding
      a random offset drawn from some symmetrical distribution to the
      current value of $x_1$.
      Keep the values of $x^{(1)}_2,\ldots,x^{(K)}_2$
      unchanged in the proposed state.
\item[\bf ~~~fast variables shifted]~\\[6pt] 
      Propose a new value for $x_1$ in all ensemble members as above,
      together
      with new values for $x^{(1)}_2,\ldots,x^{(K)}_2$ found by
      adding an offset to all of $x^{(1)}_2,\ldots,x^{(K)}_2$ that 
      is randomly drawn from some symmetrical distribution.\vspace{-6pt}
\end{description}
These two schemes are illustrated in Figure~\ref{fig-ens-up}.
Either scheme could be combined with any of the three
types of ensembles above, but shifting the fast variables when they are
sampled independently from some distribution seems unlikely to work
well, since after shifting, the values for the fast variables would
not be typical of the distribution.

One could perform just one Metropolis update on the ensemble state
before returning to a single point in $\X$, or many such updates could
be performed.  Moving back to a single point and then regenerating an
ensemble before performing the next ensemble update will requires some
computation time, but may improve sampling.  Updates might also be
performed on $\X$ (perhaps changing only fast variables) before moving
back to an ensemble.

\subsection*{Fast and slow variables in Gaussian process regression models}
\vspace*{-8pt}

Gaussian process regression (Neal 1999; Rasmussen and Williams, 2006)
is one context where ensemble MCMC can be applied to facilitate
sampling when there are ``fast'' and ``slow'' variables.

\paragraph{A brief introduction to Gaussian process regression.}
Suppose that we observe pairs $(z^{(i)},y^{(i)})$ for $i=1,\ldots,n$, where
$z^{(i)}$ is a vector of $p$ covariates, whose distribution is not
modeled, and $y^{(i)}$ the corresponding real-valued response, which
we model as $y^{(i)} \,=\, f(z^{(i)})\, +\, e^{(i)}$, where $f$ is an
unknown function, and $e^{(i)}$ is independent Gaussian noise with
mean zero and variance $\sigma^2$.  We can give $f$ a Gaussian process
prior, with mean zero and some covariance function
$C(z,z')\,=\,E[f(z)f(z')]$.  We then base prediction for a future observed
response, $y^*$, associated with covariate vector $z^*$, on the
conditional distribution of $y^*$ given $y =
(y^{(1)},\ldots,y^{(n)})$.  This conditional distribution is Gaussian,
with mean and variance that can be computed as follows:
\beq
 E(y^*|y) \ \ =\ \ k\T \Sigma^{-1} y,\ \ \ \
 \mbox{Var}(y^*|y) \ \ =\ \ v\ -\ k\T \Sigma^{-1} k
\label{eq-gp-pred}
\eeq
Here, $\Sigma = K+\sigma^2I$ is the covariance matrix of $y$, with $K_{ij} =
C(z^{(i)},z^{(j)})$.  Covariances between $y^*$ and
$y$ are given by the vector $k$, with $k_i=C(z^*,z^{(i)})$. 
The marginal variance of $y^*$ is $v=C(z^*,z^*)+\sigma^2$.

If the covariance function $C$ and the noise variance $\sigma^2$ are
known, the matrix operations above are all that are needed for inference
and prediction.  However, in practice, the noise variance is not
known, and the covariance function will also have unknown parameters.
For example, one commonly-used covariance function has the form\vspace*{-6pt}
\beq
   C(z,z') & = & \eta^2\, \Big( a^2 \ +\ \exp\Big(\!-\sum_{h=1}^p \Big(
     \nu_h\, (z_h - z_h') \Big)^2\ \Big) \Big)
   \label{eq-cov}
\eeq
Here, the $a^2$ term allows for the overall level of the function to
be shifted upwards or downwards from zero; it might be fixed \textit{a
priori}.  However, we typically do not know what are suitable values
for the parameter $\eta$, which expresses the magnitude of variation
in the function, or for the parameters $\nu_1,\ldots,\nu_p$, which
express how relevant each covariate is to predicting the response
(with $\nu_h=0$ indicating that $z_h$ has no effect on predictions for $y$).

In a fully Bayesian treatment of this model, $\eta$, $\nu$, and
$\sigma$ are given prior distributions.  Independent Gaussian prior
distributions for the log of each parameter may be suitable, with
means and variances that are fixed \textit{a priori}.  In the models I
fit below, the logs of the components of $\nu$ are correlated, which
is appropriate when learning that one covariate is highly relevant to
predicting $y$ increases our belief that other covariates might also
be relevant.  After sampling the posterior distribution of the
parameters with some MCMC method, predictions for future observations
can be made by averaging the Gaussian distributions given
by~\eqref{eq-gp-pred} over values of the parameters taken from the
MCMC run.

Sampling from the posterior distribution requires the likelihood for
the parameters given the data, which is simply the Gaussian
probability density for $y$ with these parameters.  The log
likelihood, omitting constant terms, can therefore be written as
\beq
  \log L(\eta,\nu,\sigma) & = & -(1/2)\log \det \Sigma(\eta,\nu,\sigma)
   \ -\ (1/2) y\T \Sigma(\eta,\nu,\sigma)^{-1} y
  \label{eq-gp-ll}
\eeq
This log likelihood is usually computed by finding the Cholesky
decomposition of $\Sigma$, which is the
lower-triangular matrix $L$ for which $LL\T = \Sigma$.  From $L$, both
terms above are easily calculated.  In the first term, the determinant
of $\Sigma$ can be found as $(\det L)^2$, with $\det L$ being just the 
product of the diagonal entries of $L$.  In the second term, $y\T\Sigma^{-1}y
= y\T(LL\T)^{-1}y = u\T u$, where $u = L^{-1}y$ can be found by forward
substitution.

Computing the Cholesky decomposition of $\Sigma$ requires time
proportional to $n^3$, with a rather small constant of
proportionality.  Computation of $\Sigma$ requires time proportional
to $n^2p$, with a somewhat larger constant of proportionality (which
may be larger still when covariance functions more complex than
equation~\eqref{eq-cov} are used).  If $p$ is fixed, the time for the
Cholesky decomposition will dominate for large $n$, but for moderate
$n$ (eg, $n=100$, $p=10$), the time to compute $\Sigma$ may be greater.

\paragraph{Expressing computations in a form with fast variables.}

I will show here how the likelihood for a Gaussian process regression
model can be rewritten so that some of the unknown parameters are
``fast''.  Specifically, if computations are done using the Cholesky
decomposition, $\eta$ can be a fast parameter, and if computations are
instead done by finding the eigenvectors and eigenvalues of the
covariance matrix, both $\eta$ and $\sigma$ can be fast parameters.
Note that the MCMC methods used work better when applied to the logs
of the parameters, so the actual fast parameters will be $\log(\eta)$
and $\log(\sigma)$, though I sometimes omit the logs when referring to
them here.

The form of covariance function shown in equation~\eqref{eq-cov} has
already been designed so that $\eta$ can be a fast parameter.
Previously, I (and others) have used covariance functions in which the
constant term $a$ is not multiplied by the scale factor $\eta^2$, but
writing it as in equation~\eqref{eq-cov} will be essential for $\eta$
to be a fast parameter.  As an expression of prior beliefs, it makes
little difference whether or not $a^2$ is multiplied by $\eta^2$, since $a$
is typically chosen to be large enough that any reasonable shift of
the function is possible, even if $a$ is multiplied by a value of
$\eta$ at the low end of its prior range.  Indeed, one could let $a$
go to infinity --- analogous to letting the prior for the intercept
term in an ordinary linear regression model be improper --- without
causing any significant statistical problem, although in practice, to
avoid numerical problems from near-singularity of $\Sigma$,
excessively large values of $a$ should be avoided.  A large value for
$a$ will be usually be unnecessary if the responses, $y$, are centred
to have sample mean zero.

To make $\eta$ a fast variable when using Cholesky computations, we also
need to rewrite the contribution of the noise variance to $\Sigma$ 
as $\eta^2$ times $\sigma^2/\eta^2 \,=\, \psi^2$.  We will then 
do MCMC with $\log(\eta)$ as a fast variable and $\log(\psi)$ and
$\log(\nu_h)$ for $h=1,\ldots,p$ as slow variables.  (Note that the
Jacobian for the transformation from $(\log(\eta),\log(\sigma))$ to
$(\log(\eta),\log(\psi))$ has determinant one, so no adjustment 
of posterior densities is required with this transformation.)

Fast recomputation of the likelihood after $\eta$ changes can
be done if we write $\Sigma \,=\, \eta^2 (\Upsilon + \psi^2 I)$, 
where\vspace{-10pt}
\beq
   \Upsilon_{ij} & = & a^2 \ +\ \exp\Big(\!-\sum_{h=1}^p \Big(
     \nu_h\, (z_h^{(i)} - z_h^{(j)}) \Big)^2\Big)
\eeq
$\Upsilon + \psi^2 I$ is not a function of $\log(\eta)$, but only
of $\log(\psi)$ and the $\log(\nu_h)$.  Given values for these slow
variables, we can compute the Cholesky decomposition of
$\Upsilon+\psi^2 I$, and from that $\det (\Upsilon+\psi^2 I)$ and $y\T
(\Upsilon+\psi^2 I) y$, as described earlier.  If these results are
saved, for any value of the fast variable, $\log(\eta)$, we can quickly
compute $\det \Sigma \,=\, \eta^{2n} \det (\Upsilon+\psi^2)$ and $y\T
\Sigma^{-1} y \,=\, y\T (\Upsilon+\psi^2I)^{-1} y\,/\,\eta^2$, 
which suffice for computing the likelihood.

Rather than use the Cholesky decomposition, we could instead compute
the likelihood for a Gaussian process model by finding the
eigenvectors and eigenvalues of $\Sigma$.  Let $E$ be the matrix with
the eigenvectors (of unit length) as columns, and let
$\lambda_1,\ldots,\lambda_n$ be the corresponding eigenvalues.  The
projections of the data on the eigenvectors are given by $u = E\T y$.
The log likelihood of~\eqref{eq-gp-ll} can then be computed as
follows:\vspace{-8pt}
\beq
  \log L & = & 
   -{1\over2} \sum_{i=1}^n \log \lambda_i 
   \ -\ {1\over2} \sum_{i=1}^n {u_i^2\over\lambda_i}
\eeq
Both finding the Cholesky decomposition of an $n \times n$ matrix and
finding its eigenvectors and eigenvalues take time proportional to
$n^3$, but the constant factor for finding the Cholesky decomposition
is smaller than for finding the eigenvectors (by about a factor of
fifteen), hence the usual preference for using the Cholesky decomposition.

However, if computations are done using eigenvectors and eigenvalues,
both $\eta$ and $\sigma$ can be made fast variables.  To do this, we
write $\Sigma \,=\, \eta^2 \Upsilon + \sigma^2 I$.  Given values for
the slow variables, $\log(\nu_h)$ for $h=1,\ldots,p$, we can compute
$\Upsilon$, and then find its eigenvalues,
$\lambda_1,\ldots,\lambda_n$, and the projections of $y$ on each of
its eigenvectors, which will be denoted $u_i$ for $i=1,\ldots,n$.  If
we save these $\lambda_i$ and $u_i$, we can quickly compute $L$ for
any values of $\eta$ and $\sigma$.  The eigenvectors of $\Sigma$ are
the same as those of $\Upsilon$, and the eigenvalues of $\Sigma$
are $\eta^2 \lambda_i + \sigma^2$ for $i=1,\ldots,n$.  The log likelihood
for the $\log (\nu_h)$ values used to compute $\Upsilon$ along with
values for the fast variables $\log(\eta)$ and $\log(\sigma)$ can therefore
be found as
\beq
  \log L & = & 
   -{1\over2} \sum_{i=1}^n \log\, (\eta^2 \lambda_i + \sigma^2) 
   \ -\ {1\over2} \sum_{i=1}^n {u_i^2 \over \eta^2\lambda_i+\sigma^2}
\eeq

This procedure must be modified to avoid numerical difficulties when
$\Upsilon$ is nearly singular, as can easily happen in practice.  The
fix is to add a small amount to the diagonal of $\Upsilon$, giving
$\Upsilon' \,=\ \Upsilon + r^2I$, where $r^2$ is a small amount of
``jitter'', and then using $\Upsilon'$ in the computations described
above.  The statistical effect of this is that the noise variance is
changed to $\eta^2 r^2 + \sigma^2$.  By fixing $r$ to a suitably small
value, the difference of this from $\sigma^2$ can be made negligible.
For consistency, I also use $\Upsilon'$ for the Cholesky method
(so the Cholesky decomposition is of $\Upsilon'+\psi^2 I$), even though
adding jitter is necessary in that context only when $\psi^2$ might be
very small.

To summarize, if we use the Cholesky decomposition, we can compute the
likelihood for all members of an ensemble differing only in $\eta$ using
only slightly more time than is needed to compute the likelihood for
one point in the usual way.  Only a small amount of additional time,
independent of $n$ and $p$, is needed for each member of the ensemble.
Computation of eigenvectors is about fifteen times slower than the
Cholesky decomposition, and computing the likelihood for each ensemble
member using these eigenvectors is also slower, taking time
proportional to $n$.  However, when using eigenvector computations,
both $\eta$ and $\sigma$ can be fast variables.  Which method will be
better in practice is unclear, and likely depends on the values of $n$
and $p$ --- for moderate $n$ and large $p$, time to compute the
covariance matrix, which is the same for both methods, will dominate,
favouring eigenvector computations that let $\sigma$ be a fast
variable, whereas for large $n$ and small $p$, the smaller time to
compute the Cholesky decomposition may be the dominant consideration.

If we wish to use a small ensemble over both $\eta$ and $\sigma$, it
may sometimes be desirable to do the computations using the Cholesky
decomposition, recomputing it for every value of the fast variables.
This would make sense only if the time to compute the covariances
(apart from a final scaling by $\eta^2$ and addition of $\sigma^2 I$)
dominates the time for these multiple Cholesky computations (otherwise
using the ensemble will not be beneficial), and the ensemble has less
than about fifteen members (otherwise a single eigenvector computation
would be faster).  However, I do not consider this possibility in
the demonstraton below, where larger ensembles are used.

\subsection*{A demonstration}\vspace*{-8pt}

Here, I demonstrate and compare standard Metropolis and ensemble MCMC
on the problem of sampling from the posterior distribution for a
Gaussian process regression model, using the fast-slow methods
described above.  The model and synthetic test data that I use were
chosen to produce an interesting posterior distribution, having
several modes that correspond to different degress of predictability
of the response (and hence different values for $\sigma$).  

The MCMC programs were written in R, and run on two machines, with two
versions of R --- version 2.8.1 on a Solaris/SPARC machine, and
version 2.12.0 on a Linux/Intel machine.  The programs used are
available at my web site.\footnote{
\texttt{http://www.cs.utoronto.ca/$\sim$radford/}}

\paragraph{The synthetic data.}
In the generated data, the true relationship of the response, $y$, to
the vector of covariates, $z$, was $y \,=\, f(z) + e$, with $e$ being
independent Gaussian noise with standard deviation $0.4$, and the
regression function being
\beq
   f(z) & = & 0.7z_1^2 \ +\ 0.8\sin(0.3+(4.5+0.5z_1)\,z_2)\ +\ 
              0.85\cos(0.1+5z_3+0.1z_2^2)
\eeq

The number of covariates was $p=12$, though as seen above, $f(z)$
depends only on $z_1$, $z_2$, and $z_3$.  The covariate values were
randomly drawn from a multivariate Gaussian distribution in which all
the $z_h$ had mean zero and variance one.  There were weak
correlations among $z_1$, $z_2$, and $z_3$.  There were strong (0.99)
correlations between $z_1$ and $z_4$, $z_2$ and $z_5$, and $z_3$ and
$z_6$, and moderate (0.9) correlatons between $z_1$ and $z_7$, $z_2$
and $z_8$, and $z_3$ and $z_9$.  Covariates $z_{10}$, $z_{11}$, and
$z_{12}$ were independent of each other and the other covariates.\footnote{
In detail, the covariates were generated by letting $w_h$ for $h=1,\ldots,12$
be independent standard normals, and then letting $z_1=w_1$, 
$z_2=0.25z_1+w_2\sqrt{1-0.25^2}$,
$z_3=0.25z_2+w_3\sqrt{1-0.25^2}$, 
$z_4=0.99z_1+w_4\sqrt{1-0.99^2}$,
$z_5=0.99z_2+w_5\sqrt{1-0.99^2}$,
$z_6=0.99z_3+w_6\sqrt{1-0.99^2}$,
$z_7=0.9z_1+w_7\sqrt{1-0.9^2}$,
$z_8=0.9z_2+w_8\sqrt{1-0.9^2}$,
$z_9=0.9z_3+w_9\sqrt{1-0.9^2}$,
$z_{10}=w_{10}$, $z_{11}=w_{11}$, and $z_{12}=w_{12}$.
}

I generated $n=100$ independent pairs of covariate vectors and
response values in this way.

\paragraph{The Gaussian process model.}
This synthetic data was fit with a Gaussian process regression model of
the sort described in the previous section, in which the prior covariance 
between responses $y^{(i)}$ and $y^{(j)}$ was\vspace{-6pt}
\beq
  \mbox{Cov}(y^{(i)},y^{(j)}) & = &
     \eta^2\, \Big( a^2 \ +\ \exp\Big(\!-\sum_{h=1}^p \Big(
     \nu_h\, (z_h^{(i)} - z_h^{(j)}) \Big)^2\ \Big) + r^2 \delta_{ij}\Big)
     + \sigma^2 \delta_{ij}
\eeq
where $\delta_{ij}$ is zero if $i \ne j$ and one if $i=j$.  I fixed
$a=1$ and $r=0.01$.  

The priors for $\eta$, $\sigma$, and $\nu$ were independent.  The
prior for $\log(\sigma)$ was Gaussian with mean $\log(0.5)$ and
standard deviation 1.5; that for $\log(\eta)$ was Gaussian with mean
of $\log(1)$ and standard deviation 1.5.  The prior for $\nu$ was
multivariate Gaussian with each $\log(\nu_h)$ having mean $\log(0.5)$
and standard deviation 1.8, and with the correlation of $\nu_h$ and
$\nu_{h'}$ for $h \ne h'$ being $0.69$.  Having a non-zero prior
correlation between the $\nu_h$ is equivalent to using a prior expressed
in terms of a higher-level hyperparameter, $\nu_0$, conditional on
which the $\nu_h$ are independent with mean $\nu_0$.

\paragraph{Performance of standard Metropolis sampling.}

I tried sampling from the posterior distribution for this model and
data using standard random-walk Metropolis methods, with the state
variables being the logs of the parameters of the covariance function
(the twelve $\nu_h$ parameters, $\eta$, and $\sigma$).

I first tried using a multivariate Gaussian Metropolis proposal distribution,
centred at the current state, with independent changes for each
variable, with the same standard deviation --- ie, the proposal
distribution was $N(x,s^2I)$, where $x$ is the current state, and $s$
is the proposal standard deviation.

\begin{figure}[t]

\vspace*{1in}

$s=0.25$:\vspace*{-1.6in}

\hspace*{0.6in}\psfig{file=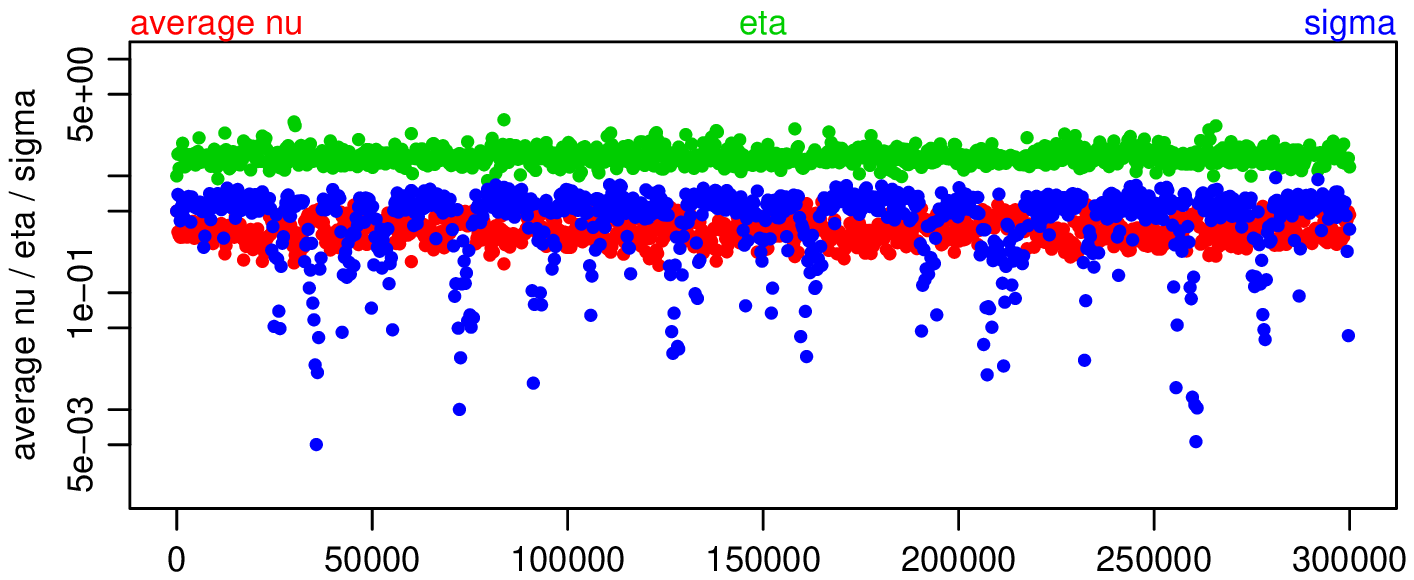}

\vspace*{0.7in}

$s=0.35$:\vspace*{-1.6in}

\hspace*{0.6in}\psfig{file=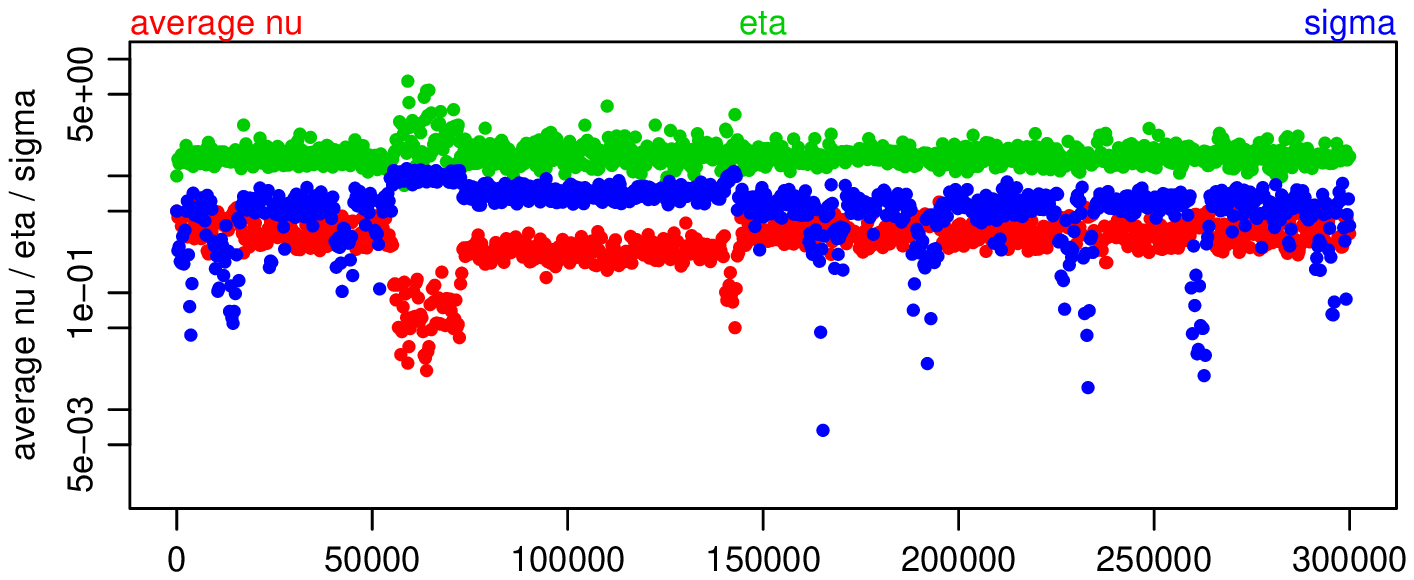}

\vspace*{-37pt}

\caption[]{Runs using Metropolis updates of all parameters at once,
with two values for the proposal standard deviation, $s$.}\label{fig-std-met-m}

\end{figure}

Figure~\ref{fig-std-met-m} shows results with $s=0.25$ (top) and
$s=0.35$ (bottom).  The plots show three quantities, on a logarithmic
scale, for every 300th iteration from a total of 300,000 Metropolis
updates (so that 1000 points are plotted).  The quantity shown in red
is the average of the $\nu_h$ parameters (giving relevances of the
covariates), that shown in green is the $\eta$ parameter (the overall
scale), and that shown in blue is the $\sigma$ parameter (noise
standard deviation).

When $s=0.25$, the rejection rate is 87\%.  From the top plot, it
appears that the chain has converged, and is sampling from two modes,
characterized by values of $\sigma$ around 0.5, or much smaller values.
Movement between these modes occurs fairly infrequently, but an
adequate sample appears to have been obtained in 300,000 iterations.

However, the bottom plot shows that this is not the case.  The
rejection rate here, with $s=0.35$, is high, at 94\%, but this chain
explores additional modes that were missed in the run with $s=0.25$.
The sample from 300,000 iterations is nevertheless far from adequate.
Some modes were moved to and from only once in the run, so an accurate estimate
of the probality of each mode cannot be obtained.

Metropolis updates that change only one parameter at a time,
systematically updating all parameters once per MCMC iteration,
perform better.  (This is presumably because in the posterior
distribution there is fairly low dependence of some parameters on
others, so that fairly large changes can sometimes be made to a single
parameter.)  Updating a $\nu_h$ parameter is slow, requiring about the
same computation time as an update changing all parameters.  However,
updates that change only $\eta$, or (if computations are done using
eigenvectors) only $\sigma$, will be fast.  To allow a fair
comparison, the number of iterations done was set so that the number
of slow evaluations was 300,000, the same as for the
runs with Metropolis updates of all variables at once.  To make
the comparison as favourable as possible for this method, I assumed
that computations are done using eigenvectors, so that both $\eta$
and $\sigma$ are fast variables.

\begin{figure}[t]

\vspace*{1.1in}

~\vspace*{-1.6in}

\hspace*{0.6in}\psfig{file=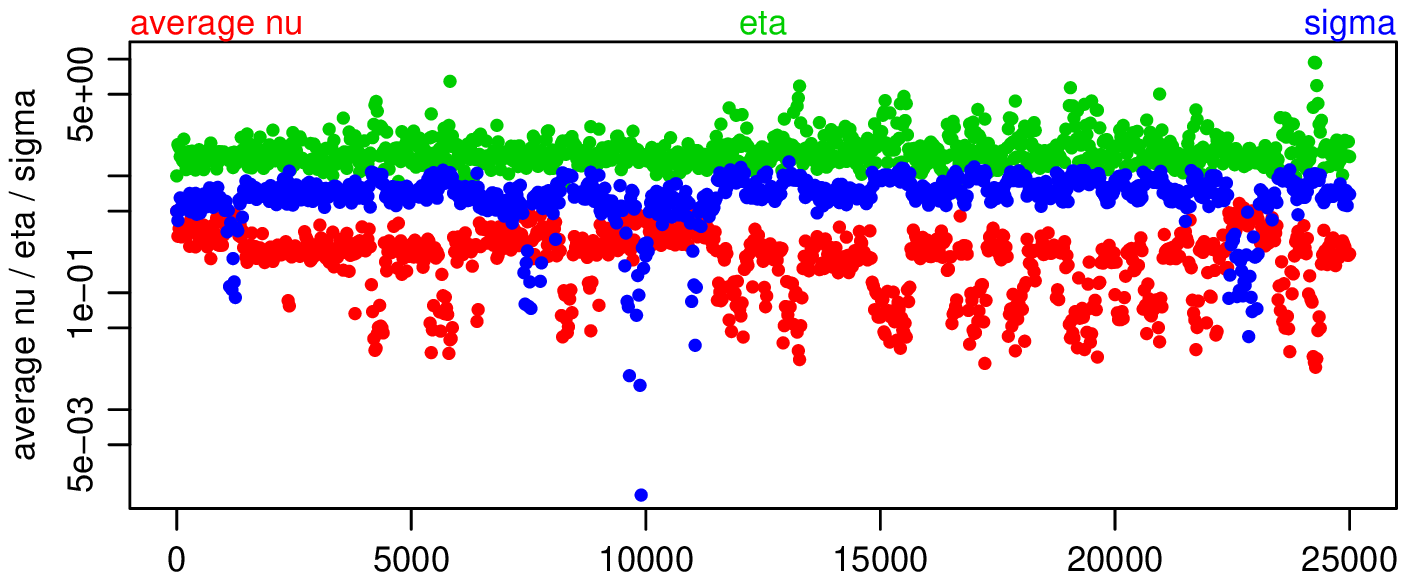}

\vspace*{0.6in}

Extra:\vspace*{-1.5in}

\hspace*{0.6in}\psfig{file=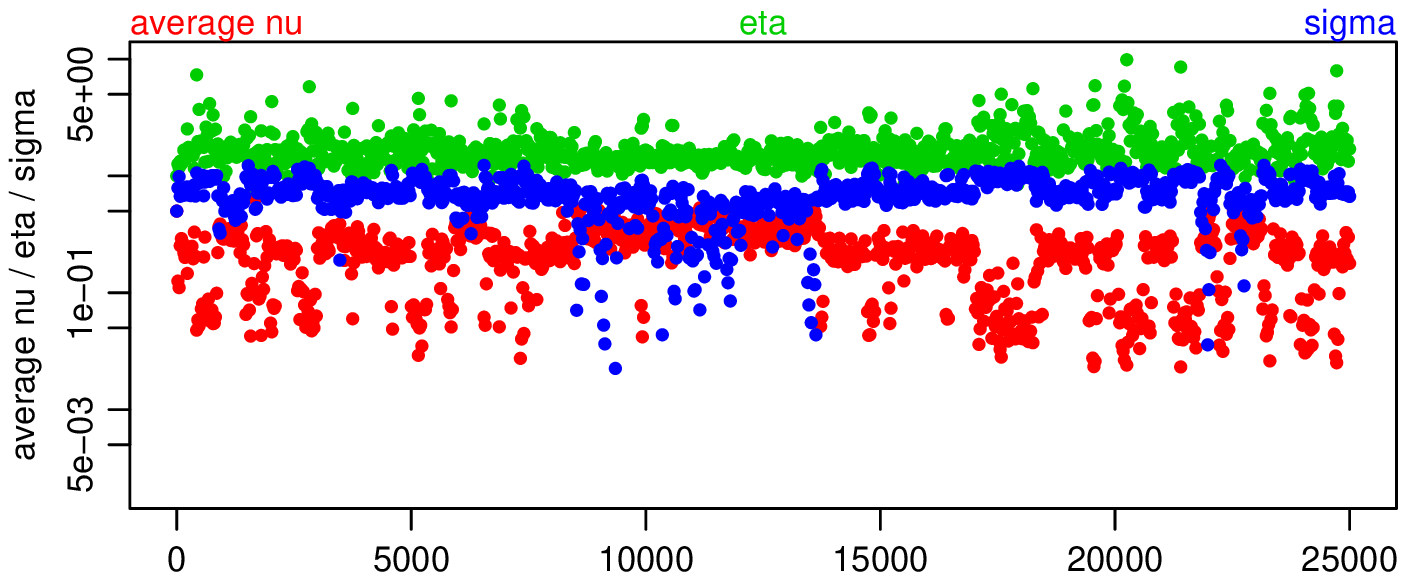}

\vspace*{-37pt}

\caption[]{Runs using Metropolis updates of one parameter at a time, with
extra updates of fast parameters for the lower plot.}\label{fig-std-met-o}

\end{figure}

The top plot in Figure~\ref{fig-std-met-o} shows results using such
single-variable Metropolis updates (selected as among the best of runs
with various settings that were tried).  Here, the proposal standard
deviation was 2 for the $\nu_h$ parameters, and 0.6 for the $\eta$ and
$\sigma$ parameters.  (Recall that all parameters are represented by
their logs.)  The rejection rates for the updates of the variables
ranged from 53\% to 77\%.  One can see that sampling is significantly
improved, but that some modes are still visited only a few times
during the run.  One way to try to improve sampling further is to
perform additional updates of the fast variables.  The bottom plot in
Figure~\ref{fig-std-met-o} shows the results when 49 additional
Metropolis updates of the fast variables ($\eta$ and $\sigma$) are
done each iteration.  This seems to produce little or no improvement
on this problem.  (However, on some other problems I have tried, extra
updates of fast variables do improve sampling.)

The time required for a slow evaluation using computations based on
the Cholesky decomposition was 0.0081 seconds on the Intel machine,
and 0.0661 seconds on the SPARC machine.  When using computations
based on eigenvectors, a slow evaluation took 0.0128 seconds on the
Intel machine, and 0.0886 seconds on the SPARC machine.  The ratio of
computation time using eigenvectors to computation time using the
Cholesky decomposition was therefore 1.58 for the Intel machine and
1.34 for the SPARC machine.  Note that this ratio is much smaller than
the ratio of times to compute eigenvectors versus the Cholesky
decomposition, since times for other operations, such as computing the
covariances, are the same for the two methods, and are not negligible
in comparison when $n=100$ and $p=12$.

\paragraph{Performance of ensemble MCMC.}

I now present the results of using ensemble MCMC methods on this
problem.  The Metropolis method was used to update the ensemble, so
that a direct comparison to the results above can be made, but of
course other types of MCMC updates for the ensemble are quite
possible.  I tried Metropolis updates where the proposals both changed
the slow variables and shifted the ensemble of values for fast
variables, but keeping the ensemble of fast variables fixed seemed to
work as well or better for this problem.  Updating only one slow
variable at a time worked better than updating all at once. 

Accordingly, the results below are for ensemble updates consisting of
a sequence of single-variable Metropolis updates for each slow
variable in turn, using Gaussian proposals centred at the current
value for the variable being updated.  After each such sequence of
updates, the ensemble was mapped to a single state, and a new ensemble
was then generated.  (One could keep the same ensemble for many
updates, but for this problem that seems undesirable, considering that
regenerating the ensemble is relatively cheap.)

I will first show the results when using computations based on
eigenvectors, for which both $\eta$ and $\sigma$ are fast variables.
Three ensembles for fast variables were tried --- an independent
ensemble, with the distribution being the Gaussian prior, an
exchangeable ensemble, with the ensemble distribution being Gaussian
with standard deviations 0.4 times the prior standard deviations, and
a $7 \times 7$ grid ensemble, with grid extent chosen uniformly
between the prior standard deviation and 1.1 times the prior standard
deviation.  All ensembles had $K = 49$ members.  The number of
iterations was set so that 300,000 slow evaluations were done, to
match the runs above using standard Metropolis updates.

\begin{figure}[p]

\vspace*{1in}

Independent:\vspace*{-1.6in}

\hspace*{0.9in}\psfig{file=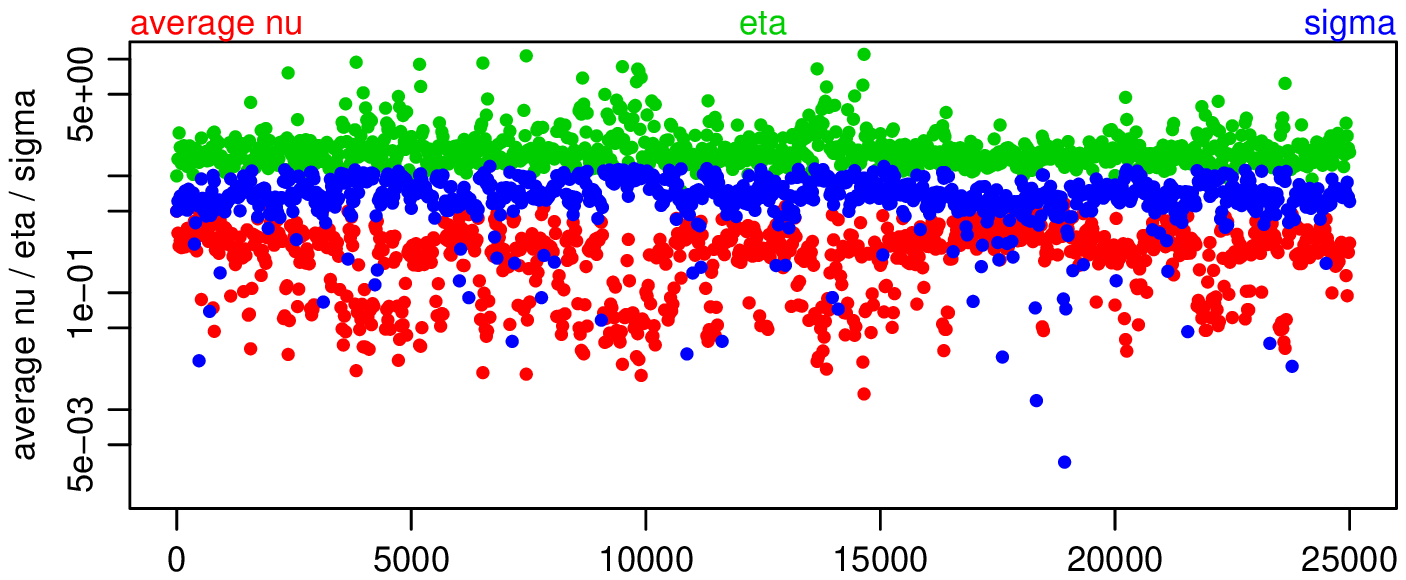}

\vspace*{0.7in}

Exchangeable:\vspace*{-1.6in}

\hspace*{0.9in}\psfig{file=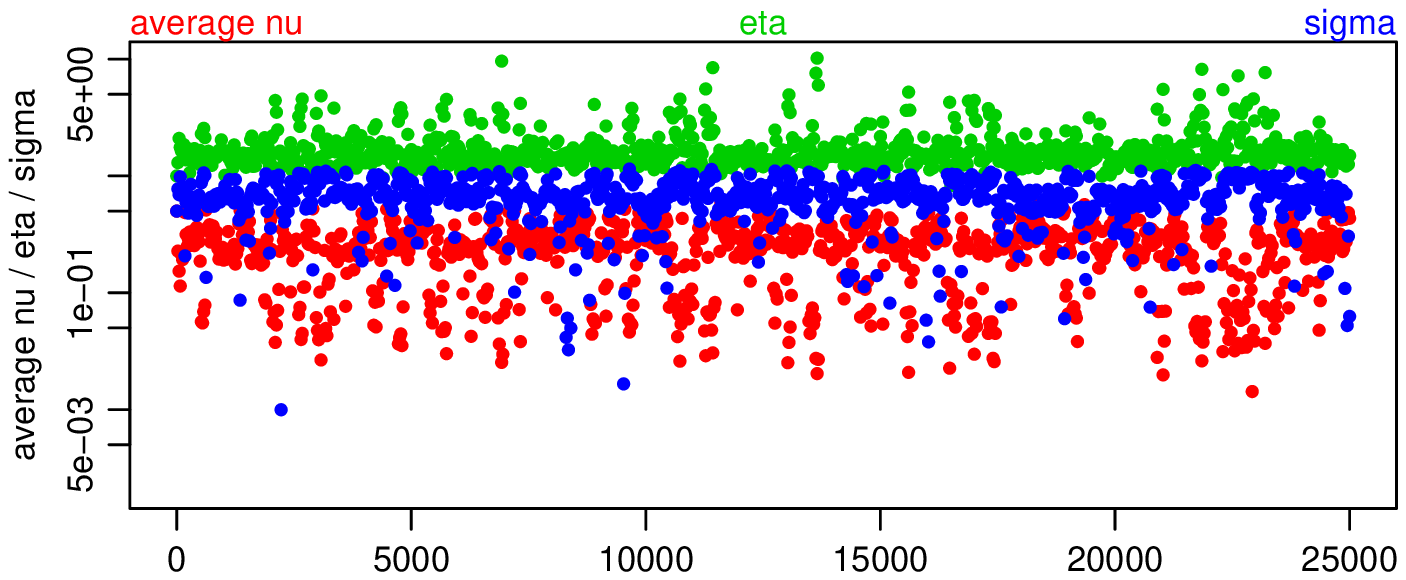}

\vspace*{0.7in}

Grid ($7 \times 7$):\vspace*{-1.6in}

\hspace*{0.9in}\psfig{file=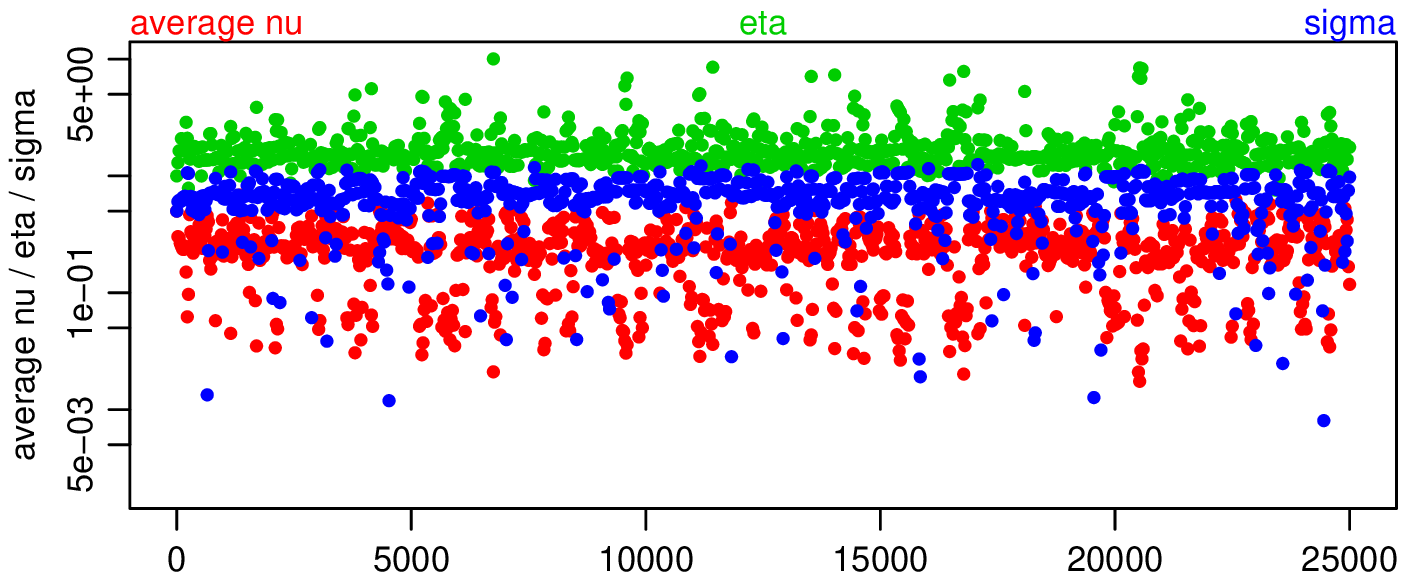}

\vspace*{-37pt}

\caption[]{Runs using ensemble MCMC with computations based on eigenvalues
(both $\eta$ and $\sigma$ fast), for three ensembles.}\label{fig-ense}

\end{figure}

Figure~\ref{fig-ense} shows the results.  Comparing with the results
of standard Metropolis in Figure~\ref{fig-std-met-o}, one can see that
there is much better movement among the modes, for all three choices
of ensemble.  The exchangeable and grid ensembles perform slightly
better than the independent ensemble.  I tried increasing the size of
the ensemble to 400, and performing extra updates of the fast
variables after mapping back to a single state, but this only
slightly improves the sampling (mostly for the independent ensemble).

The time per iteration for these ensemble methods (with $K=49$) was
0.0139 seconds on the Intel machine and 0.0936 seconds on the SPARC
machine (very close for all three ensembles).  This is slower than
standard Metropolis using Cholesky computations by a factor of 1.7 for
the Intel machine and 1.4 for the SPARC machine.  The gain in sampling
efficiency is clearly much larger than this.

% \begin{figure}[p]
% 
% \vspace*{-37pt}
% 
% \centerline{\psfig{file=rune-49-o2x-plti.ps}}
% 
% \vspace*{-57pt}
% 
% \centerline{\psfig{file=rune-49-o2x-plte.ps}}
% 
% \vspace*{-57pt}
% 
% \centerline{\psfig{file=rune-49-o2x-pltg.ps}}
% 
% \vspace*{-37pt}
% 
% \caption[]{ens eig, extra}
% \end{figure}
% 
% \begin{figure}[p]
% 
% \vspace*{-37pt}
% 
% \centerline{\psfig{file=rune-400-o2x-plti.ps}}
% 
% \vspace*{-57pt}
% 
% \centerline{\psfig{file=rune-400-o2x-plte.ps}}
% 
% \vspace*{-57pt}
% 
% \centerline{\psfig{file=rune-400-o2x-pltg.ps}}
% 
% \vspace*{-37pt}
% 
% \caption[]{ens eig, 400, extra}
% \end{figure}

\begin{figure}[p]

\vspace*{1in}

Independent:\vspace*{-1.6in}

\hspace*{0.9in}\psfig{file=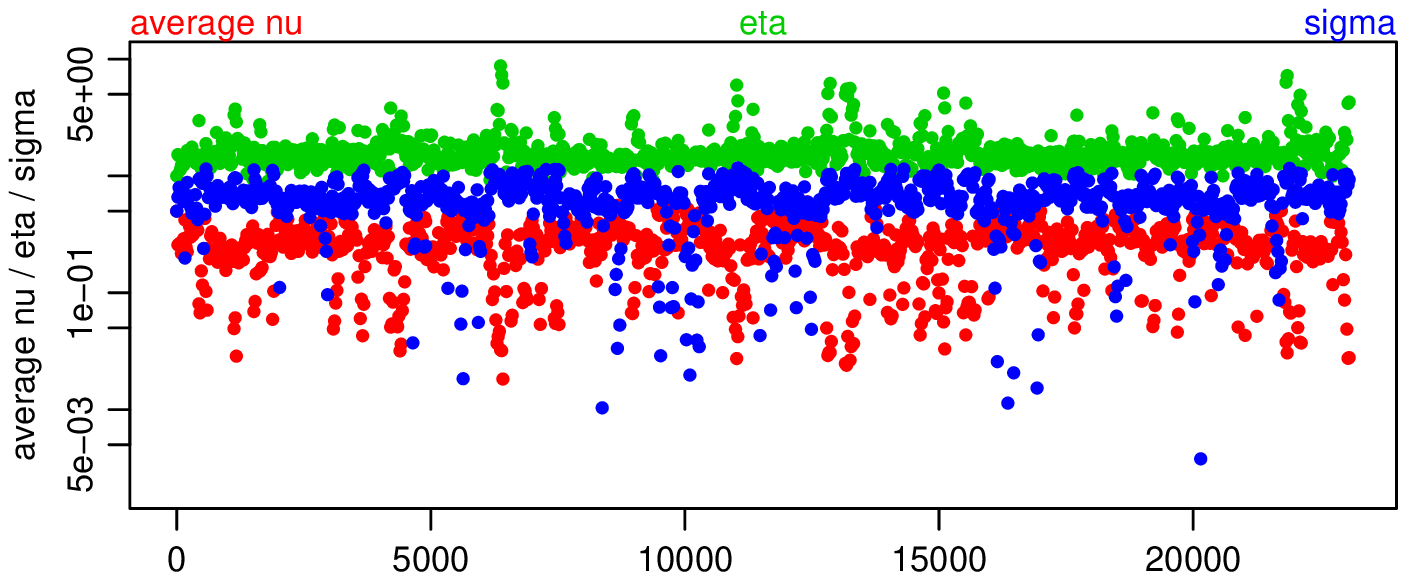}

\vspace*{0.7in}

Exchangeable:\vspace*{-1.6in}

\hspace*{0.9in}\psfig{file=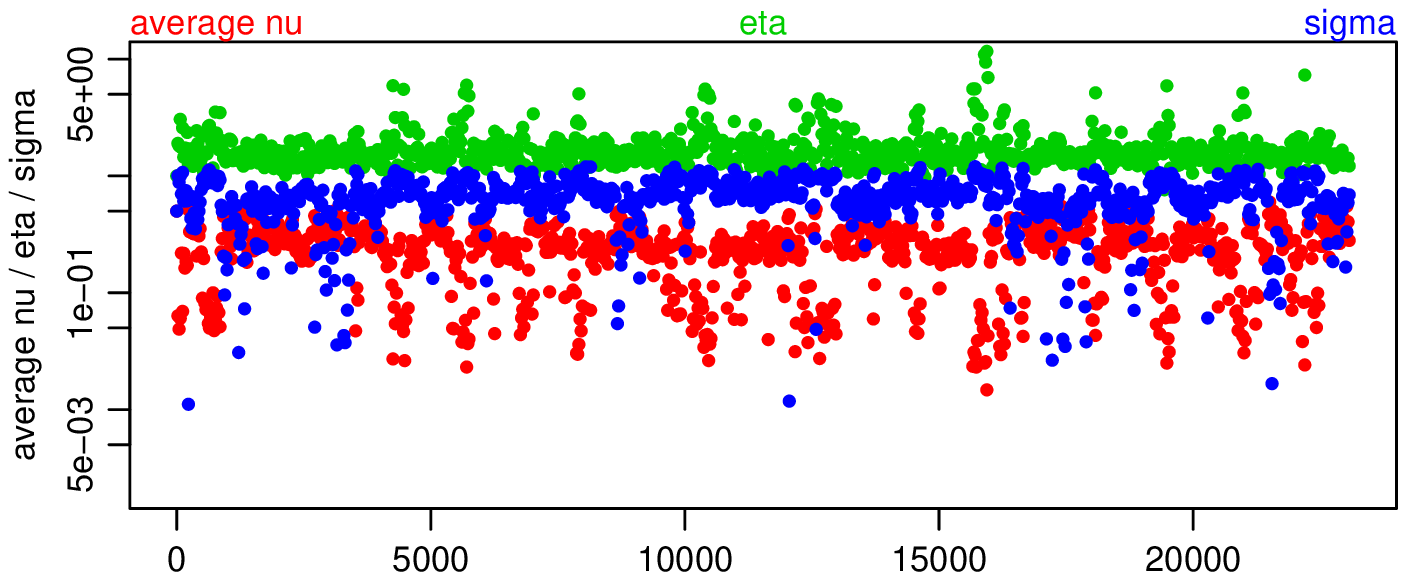}

\vspace*{0.7in}

Grid ($7 \times 7$):\vspace*{-1.6in}

\hspace*{0.9in}\psfig{file=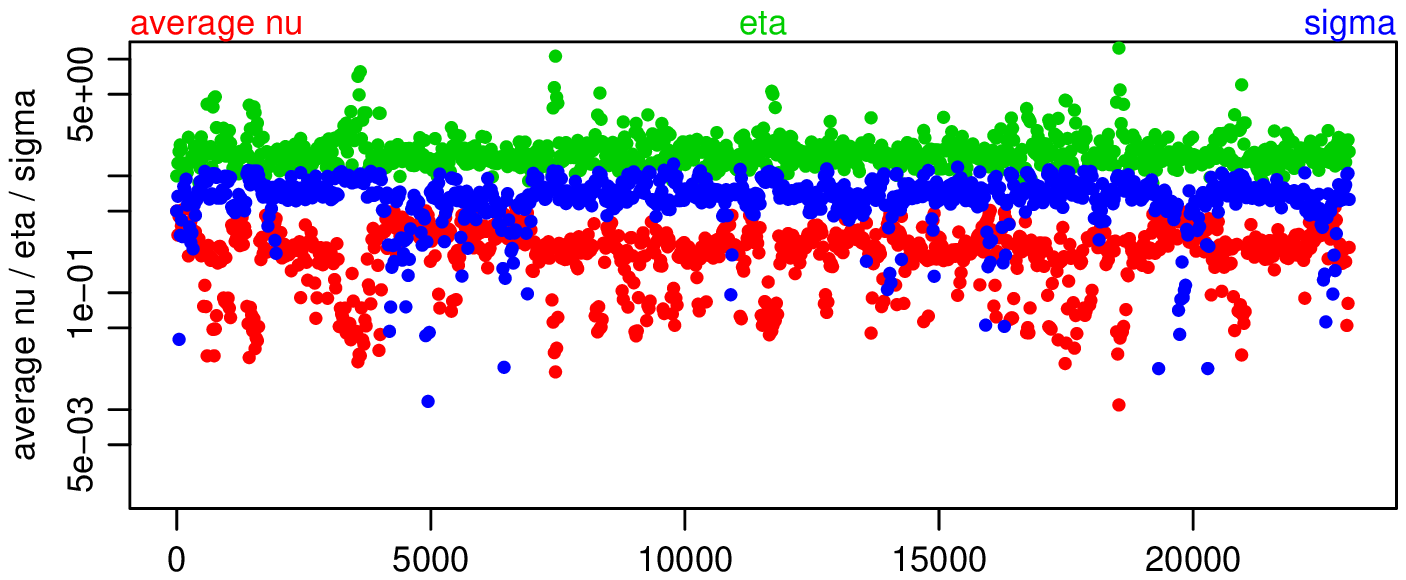}

\vspace*{-37pt}

\caption[]{Runs using ensemble MCMC with computations based on Cholesky
decomposition (only $\eta$ fast), for three ensembles.}\label{fig-ensc}

\end{figure}

Figure~\ref{fig-ensc} shows the results when computations are based on
the Cholesky decomposition, so that only $\eta$ can be a fast
variable.  Sampling is much better than with standard Metropolis, but
not as good as when both $\eta$ and $\sigma$ are fast variables.
However, the computation time is lower --- 0.0082 seconds per
iteration for on the Intel machine, and 0.0666 seconds per iteration
on the SPARC machine.  These times are only about 1\% higher than for
standard Metropolis, so use of an ensemble for $\eta$ only is
virtually free.

\subsection*{Discussion}\vspace*{-8pt}

I will conclude by discussing other possible applications of ensemble
MCMC, and its relationship with two previous MCMC methods.

\paragraph{Some other applications.}

Bayesian inference problems with fast and slow variables arise in many
contexts other than Gaussian process regression models.  As
mentioned early, many Bayesian models have hyperparameters whose
conditional distributions depend only on a fairly small number of
parameters, and which are therefore fast compared to parameters that
depend on a large set of data points.  

For example, in preliminary experiments, I have found a modest benefit
from ensemble MCMC for logistic regression models in which the prior
for regression coefficients is a $t$~distribution, whose width
parameter and degrees of freedom are hyperparameters.  When there are
$n$ observations and $p$ covariates, recomputing the posterior density
after a change only to these hyperparameters takes time proportional
to $p$, whereas recomputing the posterior density after the regression
coefficients change takes time proportional to $np$ (or to $n$, if
only one regression coefficient changes, and suitable intermediate
results were retained).  The hyperparameters may therefore be seen
as fast variables.

In Gaussian process models with latent variables, such as logistic
classification models (Neal 1999), the latent variables are all fast
compared to the parameters of the covariance function, since
recomputing the posterior density for the $n$ latent variables, given
the parameters of the covariance function, takes time proportional
$n^2$, with a small constant factor, once the Cholesky decomposition
of their covariance matrix has been found in $n^3$ time.  Looking
at ensemble MCMC for this problem would be interesting, though the
high dimensionality of the fast variables may raise additional issues.

MCMC for state-space time series models using embedded Hidden Markov
models (Neal, Beal, and Roweis, 2004) can be interpreted as an
ensemble MCMC method that maps to the ensemble and then immediately
maps back.  Looked at this way, one might consider updates to the
ensemble before mapping back.  Perhaps more promising, though, is to
use a huge ensemble of paths defined using an embedded Hidden Markov
model when updating the parameters that define the state dynamics.

\paragraph{Relationship to the multiple-try Metropolis method.}  A
Metropolis update on an ensemble of points as defined in this paper
bears some resemblence to a ``multiple-try'' Metropolis update as
defined by Liu, Liang, and Wong (2000) --- for both methods, the
update is accepted or rejected based on the ratio of two sums of terms
over $K$ points, and in certain cases, these sums are of
$\pi(x^{(i)})$ for the $K$ points, $x^{(1)},\ldots,x^{(K)}$.  In a
multiple-try Metropolis update, $K$ points are sampled independently
from some proposal distribution conditional on the current point, one
of these $K$ proposed points is then selected to be the new point if the
proposal is accepted, and a set of $K$ points is then produced by 
combining the current point with $K\!-\!1$ points sampled independently
from the proposal distribution conditional on this selected point.
Finally, whether to accept the selected point, or reject it (retaining
the current point for another iteration), is decided by a criterion
using a ratio of sums of terms for these two sets of $K$ points.

In one simple situation, multiple-try Metropolis is equivalent to
mapping from a single point to an ensemble of $K$ points, doing a
Metropolis update on this ensemble, and then mapping back to a single
point.  This equivalence arises when the proposal distribution for
multiple-try Metropolis does not actually depend on the current state,
in which case it can also be used as an independent ensemble base
distribution, and as a proposal distribution for an ensemble update in
which new values for all ensemble members are proposed
independently.\footnote{In detail, using the notation of Liu,
\textit{et al.}\ (2000), we obtain this equivalence by letting
$T(x,y)=\zeta(y)$ and $\lambda(x,y)=1/[\zeta(x)\zeta(y)]$, so that
$w(x,y)=\pi(x)T(x,y)\lambda(x,y)=\pi(x)/\zeta(x)$.} This method will
generally not be useful, however, since there is no apparent short-cut
for evaluating $\pi(x)$ at the $K$ ensemble points (or $K$ proposed
points) in less than $K$ times the computational cost of evaluating
$\pi(x)$ at one point.

Applying multiple-try Metropolis usefully to problems with fast and
slow variables, as done here for ensemble MCMC, would require that the
$K$ proposals conditional on the current state be dependent, since for
fast computation they need to all have the same values for the slow
variables.  Liu, \textit{et al.}\ mention the possibility of dependent
proposals, but provide details only for a special kind of dependence
that is not useful in this context.

In this paper I have emphasized the need for a computational short-cut
if ensemble MCMC is to provide a benefit, since otherwise it is hard
to see how an ensemble update taking a factor of $K$ more computation
time can outperform $K$ ordinary updates.  The analogous point
regarding multiple-try Metropolis was apparently not appreciated by
Liu, \textit{et al.}, as none of the examples in their paper make use
of such a short-cut.  Accordingly, in all their examples one would
expect a multiply-try Metropolis method with $K$ trials to be inferior
to simply performing $K$ ordinary Metropolis updates  with the same
proposal distribution, a comparison which is not presented in their
paper.\footnote{For their CGMC method, the $K$ ordinary Metropolis
updates should use the same reference point.  This is
valid, since the reference point is determined by a part of the state
that remains unchanged during these $K$ updates.}

\paragraph{Relationship to the multiset sampler.}  Leman, Chen, and
Lavine (2009) proposed a ``multiset sampler'' which can be seen as a
particular example of sampling from an ensemble distribution as in
this paper.  In their notation, they sample from a distribution for a
value $x$ and a multiset of $k$ values for $y$, written as
$s=\{y_1,\ldots,y_k\}$.  The $y$ values are confined to some bounded
region, $\Y$, which allows a joint density for $x$ and $s$ to be
defined as follow: \beq
   \pi^*(x,s) & \propto  & \sum_{j=1}^k \pi(x,y_j)
    \ \ =\ \ \pi(x)\, \sum_{j=1}^k \pi(y_j|x)
\eeq
where $\pi(x,y)$ is the distribution of interest.  In some examples,
they focus on the marginal $\pi(x)$.  Integrating the density above
over $y_1,\ldots,y_k$ shows that the marginal $\pi^*(x)$ is the same 
as $\pi(x)$, so that sampling $x$ and $s$ from $\pi^*$ will produces
a sample of $x$ values from $\pi$.  Leman, \textit{et al.}\ sample
from $\pi^*$ by alternating a Metropolis update for just $x$ with
with a Metropolis update for $y_j$ with $j$ selected randomly
from $\{1,\ldots,k\}$.

This $\pi^*$ distribution for the multiset sampler is the same as the
ensemble distribution that would be obtained using the methods for
fast and slow variables in this paper, if $x$ is regarded as slow
(and hence written as $x_1$ in the notation of this paper), $y$ is
regarded as fast (and hence written as $x_2$), and in the $\xi$ density
used in defining the ensemble base measure, the fast variables
in the ensemble are independent, with uniform density over $\Y$.
The density $\rho$ defined by equation~\eqref{eq-fs-rho1} then 
corresponds to the multiset density above.

Leman, \textit{et al.}\ do not distinguish between fast and slow
variables, however, and hence do not argue that the multiset sampler
would be beneficial in such a context.  Nor do they assume that there
is any other computational short-cut allowing $\pi$ to sometimes be
evaluated quickly, thereby reducing the amount of computation needed
to evaluate $\pi^*$.  They instead justify their multiset sampler as
allowing $k-1$ of the $y$ values in the multiset to freely explore the
region $\Y$, since the one remaining $y$ value can provide a
reasonably high $\pi(x,y)$ and hence also a reasonably high value for
$\pi^*$.  

The development in this paper confirms this picture.  With the $\rho$
distribution corresponding to multiset sampling, the mapping $\Tu$
will (in the notation of Leman, \textit{et al.}) go from a single
$(x,y)$ pair drawn from $\pi(x,y)$ to an ensemble with $k$ values for
$y$, one of which is the original $y$, and the other $k-1$ of which
are drawn independently and uniformly from $\Y$.  This is therefore
the equilibrium distribuiton of the multiset samplier.  The
development here also shows that $\pi(x,y)$ can be recovered by
applying the $\Td$ mapping, which will select a $y$ from the multiset
with probabilities proportional to $\pi(x,y)$.  This makes the complex
calculations in Section~6 of (Leman, \textit{et al.}, 2002)
unnecessary.

Unfortunately, it also seems that any benefit of the multiset sampler
can be obtained much more simply and efficiently with a Markov chain
sampler that randomly chooses between two Metropolis-Hastings updates
on the original $\pi$ distribution --- one that proposes a new value
for $x$ (from some suitable proposal distribution) with $y$ unchanged,
and another that proposes a new value for $x$ along with a new value
for $y$ that is drawn uniformly from $\Y$.  As is the case as well for
ensemble MCMC and multiple-try Metropolis, one should expect that looking
at $K$ points at once will produce a benefit only when a short-cut
allows $\pi$ for all these points to be computed in less than $K$
times the cost of computing $\pi$ for one point.

\subsection*{Acknowledgements}\vspace*{-8pt}

This research was supported by Natural Sciences and Engineering
Research Council of Canada. The author holds a Canada Research Chair
in Statistics and Machine Learning.

\subsection*{References}\vspace*{-8pt}

\leftmargini 0.2in
\labelsep 0in

\begin{description}
\itemsep 2pt

\item[] Earl, D.~J.\ and Deem, M.~W.\ (2005) ``Parallel tempering: Theory, 
applications, and new perspectives'', \textit{Physical Chemistry Chemical
Physics}, vol.~7, pp.~216--222.

\item[] Geyer, C.~J.\ (1991) ``Markov chain Monte Carlo maximum
likelihood'', in E.~M.~Keramidas (editor), {\em Computing Science and
Statistics: Proceedings of the 23rd Symposium on the Interface},
pp.~156-163, Interface Foundation.  Also available from
\texttt{http://www.stat.umn.edu/$\sim$charlie/}.

\item[] Gilks, W.~R., Roberts, G.~O., and George, E.~I.\ (1994)
``Adaptive direction sampling'', \textit{The Statistician (Journal of
the Royal Statistical Society, Series D)}, vol.~43, pp.~179--189.

\item[]
  Hastings, W.~K.\ (1970) ``Monte Carlo sampling methods using Markov chains 
  and their applications'', {\em Biometrika}, vol.~57, pp.~97--109.

\item[] Leman, S.~C., Chen, Y., and Lavine, M.\ (2009) ``The multiset
sampler'', \textit{Journal of the American Statistical Association},
vol.~104, pp.~1029--1041.

\item[] Lewis, A.\ and Bridle, S.\ (2002) ``Cosmological parameters
from CMB and other data: A Monte Carlo approach'', \textit{Physical
Review D}, vol.~66, 103511, 16 pages.  Also available at
\texttt{http://arxiv.org/abs/astro-ph/0205436}.  Further notes on 
MCMC methods for this problem are available at
\texttt{http://cosmologist.info/notes/CosmoMC.pdf}.

\item[] Liu, J.~S., Liang, F., and Wong, W.~H.\ (2000) ``The
multiple-try method and local optimization in Metropolis sampling'',
\textit{Journal of the American Statistical Association}, vol.~95,
pp.~121--134.

\item[]
  Metropolis, N., Rosenbluth, A.~W., Rosenbluth, M.~N., Teller, A.~H., 
  and Teller, E.\ (1953) ``Equation of state calculations by fast computing 
  machines'', {\em Journal of Chemical Physics}, vol.~21, pp.~1087--1092.

\item[] Neal, R.~M.\ (1999) ``Regression and classification using
Gaussian process priors'' (with discussion), in J.~M.~Bernardo, {\em
et al} (editors) {\em Bayesian Statistics 6}, Oxford University Press,
pp.~475--501.

\item[] Neal, R.~M.\ (2004) ``Taking bigger Metropolis steps by
dragging fast variables'', Technical Report No.\ 0411, Dept.\ of
Statistics, University of Toronto, 9 pages.

\item[] Neal, R.~M.\ (2006) ``Constructing efficient MCMC methods
using temporary mapping and caching'', talk at Columbia University,
Department of Statistics, 11 December 2006.  Slides available at
\texttt{http://www.cs.utoronto.ca/$\sim$radford/ftp/cache-map.pdf}.

\item[] Neal, R.~M.\ (2010) ``MCMC using Hamiltonian dynamics'', to
appear in the \textit{Handbook of Markov Chain Monte Carlo}, S.~Brooks,
A.~Gelman, G.~Jones, and X.-L.~Meng (editors), Chapman \& Hall / CRC
Press, 51 pages.  Also at
\texttt{http://www.cs.toronto.edu/$\sim$radford/ham-mcmc.abstract.html}.

\item[] Neal, R.~M., Beal, M.~J., and Roweis, S.~T.\ (2004) ``Inferring
state sequences for non-linear systems with embedded hidden Markov
models'', in S. Thrun, et al (editors), \textit{Advances in Neural Information
Processing Systems 16 (aka NIPS*2003)}, MIT Press, 8 pages.

\item[] Rasmussen, C.~E.\ and Williams, C.~K.~I.\ (2006)
\textit{Gaussian Processes for Machine Learning}, The MIT Press.

\end{description}

\end{document}